\documentclass{aa}
\usepackage{graphicx}
\begin{document}

\title{Magnetic fields in merging spirals -- the Antennae}

\author{Krzysztof T. Chy\.zy\inst{1}
 \and Rainer Beck\inst{2}}
\institute{Astronomical Observatory, Jagiellonian
University, ul. Orla 171, 30-244 Krak\'ow, Poland
 \and Max-Planck-Institut f\"ur Radioastronomie, Auf dem H\"ugel 69, 
53121 Bonn, Germany }
\offprints{K.T. Chy\.zy, \email{chris@oa.uj.edu.pl}}

\date{Received 26 August 2003 / Accepted 22 December 2003}

\titlerunning{Magnetic fields in the Antennae}
\authorrunning{K. Chy\.zy et al.}

\abstract{
We present an extensive study of magnetic fields in a system of 
merging galaxies. We obtained for NGC~4038/39 (the Antennae) radio 
total intensity and polarization maps at 8.44~GHz, 4.86~GHz and 1.49~GHz 
using the VLA in the C and D configurations. The galaxy pair possesses 
bright, extended radio emission filling the body of the whole system, with 
no dominant nuclear sources. The radio thermal fraction of NGC~4038/39 
was found to be about 50\% at 10.45~GHz, higher than in normal spirals. 
Most of the thermal emission is associated with star-forming regions, but 
only a part of these are weakly visible in the optical domain because of 
strong obscuration. The mean total magnetic fields in both galaxies are 
about two times stronger ($\simeq 20\,\mu$G) than in normal spirals. However, 
the degree of field regularity is rather low, implying tangling of the 
regular component in regions with interaction-enhanced star formation. 
Our data combined with those in \ion{H}{i}, H$\alpha$, X-rays and in far 
infrared allow us to study local interrelations between different gas 
phases and magnetic fields. We distinguish several radio-emitting regions 
with different physical properties and at various evolutionarystages: the 
rudimentary magnetic spiral, the northern cool part of the dark cloud complex 
extending between the galaxies, its warm southern region, its southernmost 
star-forming region deficient in radio emission, and the highly polarized 
northeastern ridge associated with the base of an unfolding tidal tail. The 
whole region of the dark cloud complex shows a coherent magnetic field 
structure, probably tracing the line of collision between the arms of merging 
spirals while the total radio emission reveals hidden star formation nests.
The southern region is a particularly intense merger-triggered starburst.
Highly tangled magnetic fields reach there strengths of $\simeq 30\,\mu$G, even 
larger than in both individual galaxies, possibly due to compression of the 
original fields pulled out from the parent disks. In the northeastern ridge, 
away from star-forming regions, the magnetic field is highly coherent with a 
strong regular component of $10\,\mu$G tracing gas shearing motions along the 
tidal tail. We find no signs of field compression by infalling gas there. 
The radio spectrum is much steeper in this region indicating aging of the CR 
electron population as they move away from their sources in star-forming 
regions. Modelling Faraday rotation data show that we deal with a 
three-dimensionally curved structure of magnetic fields, becoming almost 
parallel to the sky plane in the southeastern part of the ridge.
\keywords{Galaxies: general -- Galaxies: magnetic fields -- 
Galaxies: interactions -- Galaxies: individual: Antennae -- 
Radio continuum: galaxies}
}

\maketitle

\section{Introduction}
\label{intro}
Violently disrupted galaxies show strong departures of gas flow patterns from 
axial symmetry and from symmetric spiral shape. Encounters of gaseous disks at 
velocities in excess of 100~km/s give rise to galaxy-scale shocks and velocity 
gradients. In this case magnetic fields can provide useful information on the 
gas flows {\it in the sky plane}, complementary to any emission-line radial 
velocity studies. In particular the magnetic field can trace regions of gas 
compression (like in Virgo Cluster spirals, Chy\.zy et al. 
\cite{chyzy00}, \cite{chyzy02}) and of shearing motions stretching the 
magnetic fields. It is not yet known which gas phases have the strongest 
influence on the magnetic field evolution. While in normal spirals various gas 
phases seem to be coupled via density waves (cold gas in dust lanes, warm and 
hot ionized gas in star-forming regions in spiral arms) in violently 
interacting galaxies various gas phases appear often separated. Cold, warm and 
hot gas may occupy different disk regions. This offers an opportunity of 
studying the association of various magnetic field structures with particular 
phases of the interstellar medium. 

The degree to which the field structure follows the gas flows is unknown. 
While concepts of magnetic field generation (see e.g. Beck et al. 
\cite{beck96}, Widrow \cite{widrow}) postulate the turbulent magnetic 
diffusion coefficient to be by three orders of magnitude greater than 
that resulting from flux freezing, there are also examples of magnetic 
fields following to a large extent the large-scale gas flows (Beck et 
al. \cite{beck99}). In a strongly turbulent environment the magnetic 
field may even become insensitive to strong density wave effects 
(Soida et al. \cite{soida}).

Violently interacting galaxies also 
constitute a good tool to study the global magnetic field evolution. During 
their close encounters the timescale of changing their magnetic field 
structures is of order of $10^8$ years, much shorter than e.g. the timescale 
of classical dynamo process (e.g. Brandenburg \& Urpin \cite{brand}) but 
comparable to that of fast dynamos (Moss et al. \cite{moss}, Hanasz et al. 
\cite{hanasz}). There is a 
question to which degree external influences can amplify magnetic fields via 
the effects of compression and/or by providing a strong non-azimuthal magnetic 
fields in a manner similar to the dynamo process. To distinguish between the 
compressed, highly anisotropic fields and unidirectional ones, generated by 
the dynamo process, Faraday rotation information, thus  polarization studies 
at least at two frequencies are needed. 

\begin{figure} [t]
\resizebox{\hsize}{!}{\includegraphics{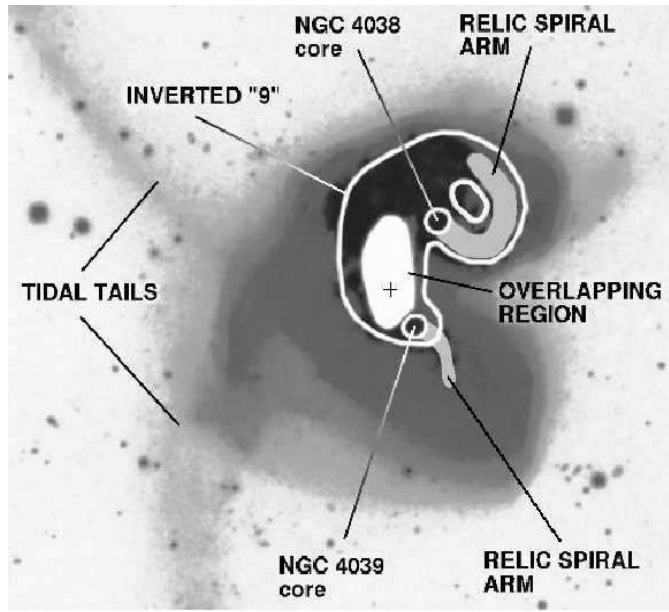}}
\caption{
Sketch of the NGC~4038/39 merging system with an enhanced DSS optical image 
in the background. The black cross marks a concentration of large complex 
of dark dusty clouds in the ``overlapping region''.
}
\label{cartoon}
\end{figure}

In this work we present a high-resolution, three-frequency radio polarization 
study of the pair of merging spirals NGC~4038/NGC~4039, called ``the 
Antennae''. This is the first study dedicated to 
magnetic fields in a system of merging 
galaxies. The galaxies are known to be strongly tidally interacting (Toomre 
\& Toomre \cite{toomre}) and forming two impressive tidal tails (see
cartoon sketch in Fig.~\ref{cartoon}). The neutral 
gas makes a long extension along the southern tail (Hibbard et al. 
\cite{hibbard}) but the northern extension is devoid of neutral gas. The 
galaxies are subject to a burst of star formation (Liang et al. \cite{liang} 
and references therein). The young star clusters are distributed along a 
peculiar structure resembling an ``inverted 9'' (Fig.~\ref{cartoon}), 
encircling the northern galaxy from the west, then running along the eastern 
edge of the whole system, 
then turning west to joining the southern member of the pair. A huge complex 
of cold dusty matter is visible in the sub-mm observations with SCUBA (Haas 
et al. \cite{haas}) in the ``overlap area'' of the galaxy disks concentrated 
 around RA$_{2000}=12^{\mathrm h} 01^{\mathrm m} 54\fs 7$ Dec$_{2000}=-18\degr 
52\arcmin 52\arcsec$ (denoted by a cross in Fig.~\ref{cartoon}). It coincides 
also with a strong CO peak found by 
Stanford et al. (\cite{stanford}). The mid-infrared ISO observations 
(Mirabel et al. \cite{mirabel}) shows that the most intense starburst occurs 
in optically obscured place to the east from the southern galaxy where 
emission comes from gas and dust heated by massive stars. The ionized gas 
in NGC~4038/39 shows a complex velocity field (Amram et al. \cite{amram}) 
while the X-ray emitting hot gas has a peculiar distribution with the 
evidence of collimated outflows (Read et al. \cite{read}). High resolution 
{\em Chandra} ACIS data (Fabbiano et al. \cite{fabbiano}) reveals a population 
of extremely luminous point-like sources (probably black hole binaries) and 
soft thermal X-ray emission associated with star-forming knots but sometimes 
also intermingle with warm (H$\alpha$) gas.
%=============================
%=============================
\begin{figure} [th]
\resizebox{\hsize}{!}{\includegraphics{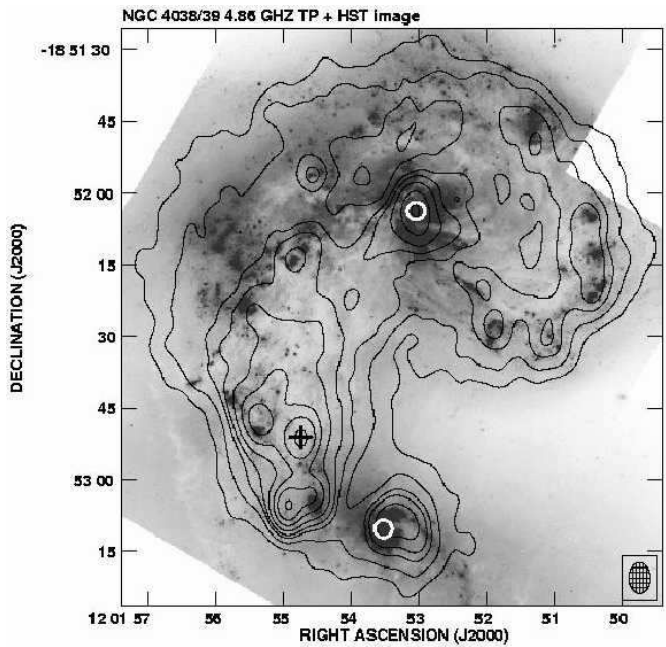}}
\caption{
Total intensity contours of radio emission of NGC~4038/39 at 4.86~GHz made from 
a combination of VLA C and D-array data overlaid upon the optical image from HST 
(courtesy B.C. Whitmore from STScI). The contour levels are 0.2, 0.4, 0.7, 
1.1, 1.7, 2.9, 4.5, 7.4 mJy/b.a. The rms noise is 0.021~mJy/b.a. and the beam 
size is $6\farcs 5\times 4\farcs 5$ (uniform weighting). White circles mark 
the positions of galaxies cores and the black cross a concentration of cold 
dust in the ``overlapping region''.
}
\label{tphig}
\end{figure}
%=============================

The galaxy pair NGC~4038/39 has been studied using VLA at 1.47~GHz and 
4.89~GHz by Hummel \& van der Hulst (\cite{hummel}). They found a radio ridge 
extending along the mentioned ``inverted 9'' structure, however, their maps do 
not show much extended structure, moreover the authors did not analyze the 
polarization information.

In this paper we present maps of NGC~4038/39 in the VLA D-configuration at 
8.44~GHz, combined C+D-array maps at 4.86~GHz as well as C-array data at 
1.49~GHz, much more sensitive to extended diffuse emission than the
earlier maps by Hummel \& van der Hulst (\cite{hummel}). 
After presenting some observational aspects of our radio data in Sect.~\ref{obs},
we introduce the total radio and polarized intensity maps in Sect.~\ref{results}
together with the distribution of spectral index, Faraday rotation and 
depolarization. Decomposition of thermal and nonthermal radio intensity
as well as polarization data results in identifying distinct radio-emitting 
regions (Sect.~\ref{discussion}). By means of profiles along different 
direction across the merging system we study the association of magnetic 
fields with different gas phases using available data in radio continuum,
 \ion{H}{i}, optical, H$\alpha$, X-ray, and dust emission (Sect.~\ref{thnth}).
In Sect.~\ref{strength} we discuss total and regular magnetic field strengths 
derived in various parts of the interacting system and compare them to values 
known from other galaxies. We give some idea of the magnetic field origin and
its regularity in Sect.~\ref{origin}. Finally, we present a tentative scenario 
of the gas and magnetic field evolution in the Antennae in Sect.~\ref{evolution}. 
The summary and suggestions of further works are given in Sect.~\ref{summary}.

\section{Observations and data reduction}
\label{obs}

Observations were made at 8.44~GHz and at 4.86~GHz, using the VLA of the 
National Radio Astronomy Observatory\footnote{NRAO is a facility of National 
Science Foundation operated under cooperative agreement by Associated 
Universities, Inc.} in the compact D-array configuration. Additional 
observations at 4.86~GHz and the measurements at 1.49~GHz were performed with 
the C-array. The observation times in C-array were 5.5 hours at 1.49~GHz and 
6 hours at 4.86~GHz, while in D-array the observation times were 5 hours at 
8.44~GHz and 5.5 hours at 4.86~GHz.

The intensity scale at all frequencies was calibrated by observing 3C286, 
adopting the flux densities of Baars et al. (\cite{baars}). The position 
angle of the linearly polarized intensity was calibrated using the same source 
with an assumed position angle of 33$\degr$. At 4.86~GHz and 8.44~GHz the 
calibrator 1157-215 was used to determine the telescope phases and the 
instrumental polarization. To check these calibration procedures 3C138 was 
observed once each day. At 1.49~GHz the phase calibrator was 1156-221. 

The data reduction has been performed using the AIPS data reduction package. 
The edited visibility data, calibrated and self-calibrated in phase, 
were Fourier transformed to obtain 
maps in Stokes parameters I, Q and U at three frequencies. At 4.86~GHz the maps 
were made using a combination of C and D-arrays with various weightings for 
particular purposes. The combined data were self-calibrated in phase.
To increase the sensitivity to extended structures we 
merged the I, Q and U data at 8.44~GHz in the UV plane with observations at 
10.45~GHz made by us with the 100-m Effelsberg radio telescope\footnote{The 
100-m telescope at Effelsberg is operated by the Max-Planck-Institut f\"ur 
Radioastronomie (MPIfR) on behalf of the Max-Planck-Gesellschaft}. The 
brightness values at 10.45~GHz were rescaled to 8.44~GHz assuming a spectral 
index of 0.7. 
Faraday rotation measures only scarcely exceed 100\,rad\,m$^{-2}$ 
(Fig.~\ref{rm}) which corresponds to a polarization angle offset of $3\degr$ 
between 10.45 and 8.44~GHz. By comparing the merged VLA/Effelsberg maps 
and pure VLA ones we state that any possible polarization angle offset between 
10.45~GHz and 8.44~GHz does not introduce visible artifacts in the merging 
process. 

Finally the Q and U maps at three frequencies were combined to get 
distributions of the linearly polarized intensity (corrected for the positive 
zero level offset) and of the position angle of the apparent magnetic vectors 
(B-vectors; not corrected for Faraday rotation observed E-vectors rotated by 
$90\degr$).

\section{Results}
\label{results}

\subsection{Total radio intensity}
\label{total1}

Details of the distribution of total radio intensity in NGC~4038/39 are 
best visible in our combined high-resolution C+D-array map at 4.86~GHz 
(Fig.~\ref{tphig}). The bright emission in the NW region follows the 
``inverted 9'' feature as already described by Hummel \& van der Hulst 
(\cite{hummel}). Local total intensity peaks coincide with the nuclear regions of 
both galaxies. An extended, radio-bright feature is located in the southern 
part of mentioned large complex of dark clouds where a mixture of dark patches 
and \ion{H}{ii} regions is present (Whitmore et al. \cite{whitmore}). While localized
peaks along the ``inverted 9'' in the NW portion coincide with particular 
star-forming complexes, the mentioned radio-bright region does not show 
particularly bright H$\alpha$-emission and its peaks seem to be loosely 
associated with optically bright regions (Fig.~\ref{tphig}). No radio emission 
was detected from the gas-poor northern tidal plume.

\begin{figure*} [t]
\begin{minipage}[b]{1\textwidth}
\centering
\includegraphics[totalheight=7.9cm]{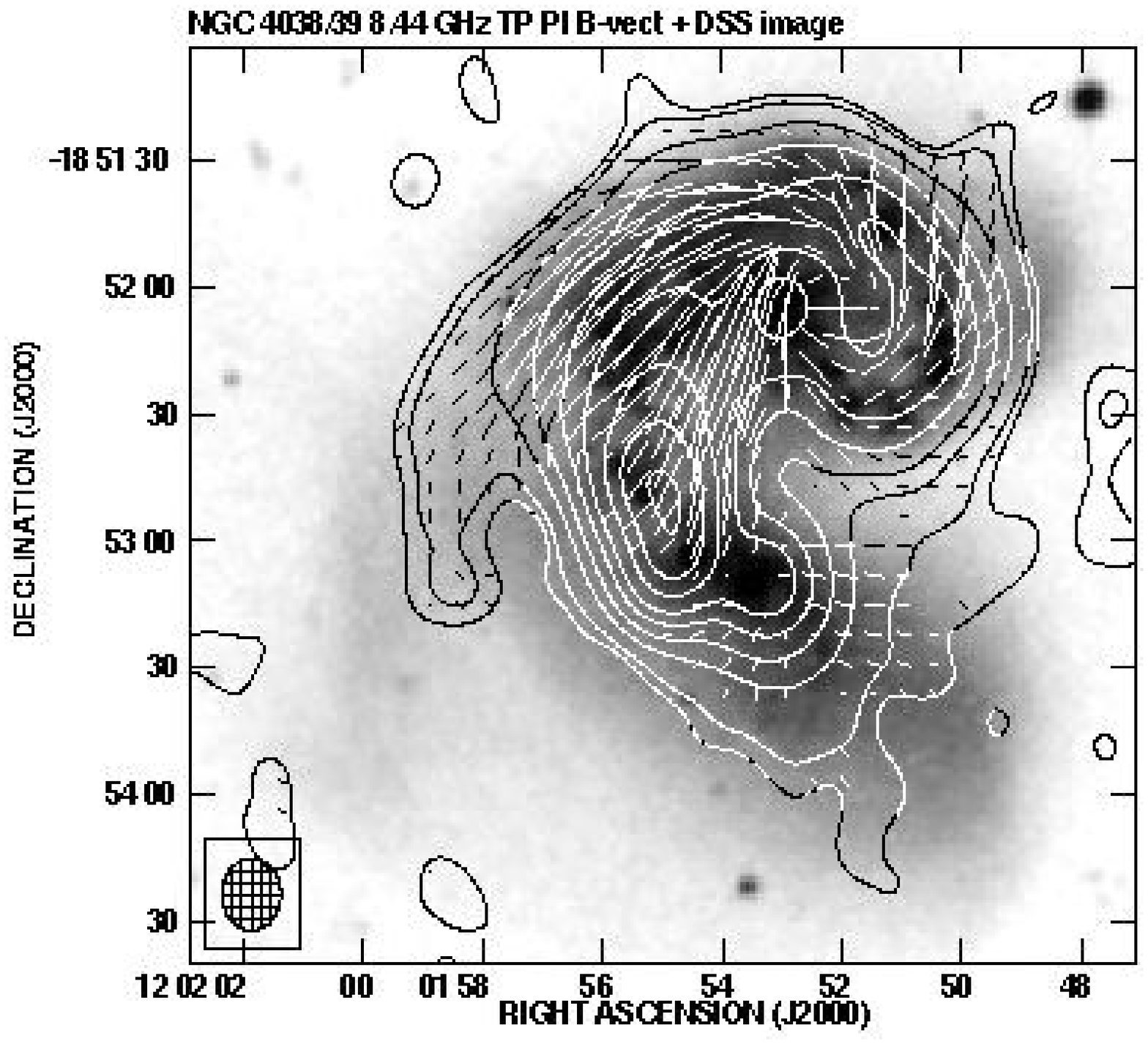}
\includegraphics[totalheight=7.9cm]{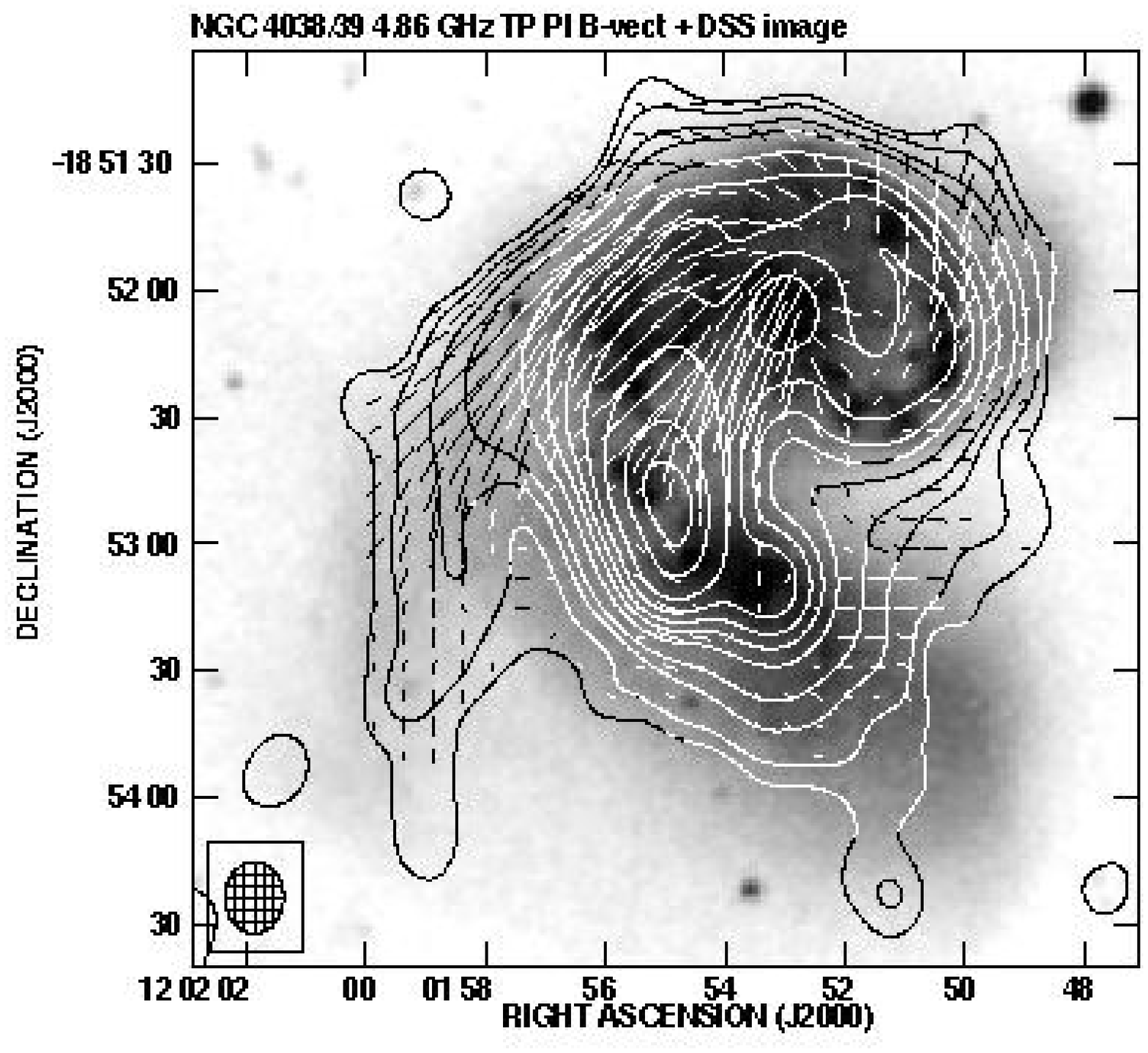}
\caption{
Left panel: the total radio intensity contours and apparent B-vectors (not 
corrected for Faraday rotation) of polarized intensity of 
NGC~4038/39 at 8.44~GHz made from a combination of VLA and Effelsberg data 
with a resolution of $17\arcsec\times 14\arcsec$ (natural weighting), 
overlaid upon a DSS image. The contour levels are 0.07, 0.11, 0.33, 0.77,
1.8, 3.3, 6.6, 11, 18 mJy/b.a., the rms noise is 0.022 mJy/b.a., a vector 
of length of $100\arcsec$ corresponds to a polarized intensity of 
$1.25~$mJy/b.a. Right panel: the contours of total intensity and apparent 
B-vectors of polarized intensity of NGC~4038/39 at 4.86~GHz made from a 
combination of VLA data in C and D with a resolution of $17\arcsec\times 
14\arcsec$ (naturally weighed and convolved) overlaid upon the DSS image. 
The contour levels are 0.05, 0.12, 0.30, 0.53, 1.2, 2.1, 3.3, 5.3, 9.0, 17,
24 mJy/b.a. and the rms noise is 0.015~mJy/b.a., a vector of length of 
$100\arcsec$ corresponds to a polarized intensity of $1.25~$mJy/b.a.}
\label{tppi}
\end{minipage}\\[30pt]
\end{figure*}

The low-resolution total intensity maps of NGC~4038/39 at 8.44~GHz and 4.86~GHz 
(Fig.~\ref{tppi}) show consistently a strong emission peak in the southern 
part of the dark cloud complex extending between the galaxies. Some weak 
maxima coincident with their nuclei are seen as well. Both maps show also an 
extension along the southern gas-rich tidal tail. No emission is detected from 
the gas-poor northern tail. A weak tongue of emission towards the southwest is 
seen consistently at both frequencies and is probably real. The total intensity
maps at both frequencies also possess a steep brightness gradient along the 
NE straight boundary of the interacting pair.

%\clearpage
%=============================

\begin{figure*} [t]
\begin{minipage}[b]{1\textwidth}
\centering
\includegraphics[totalheight=7.9cm]{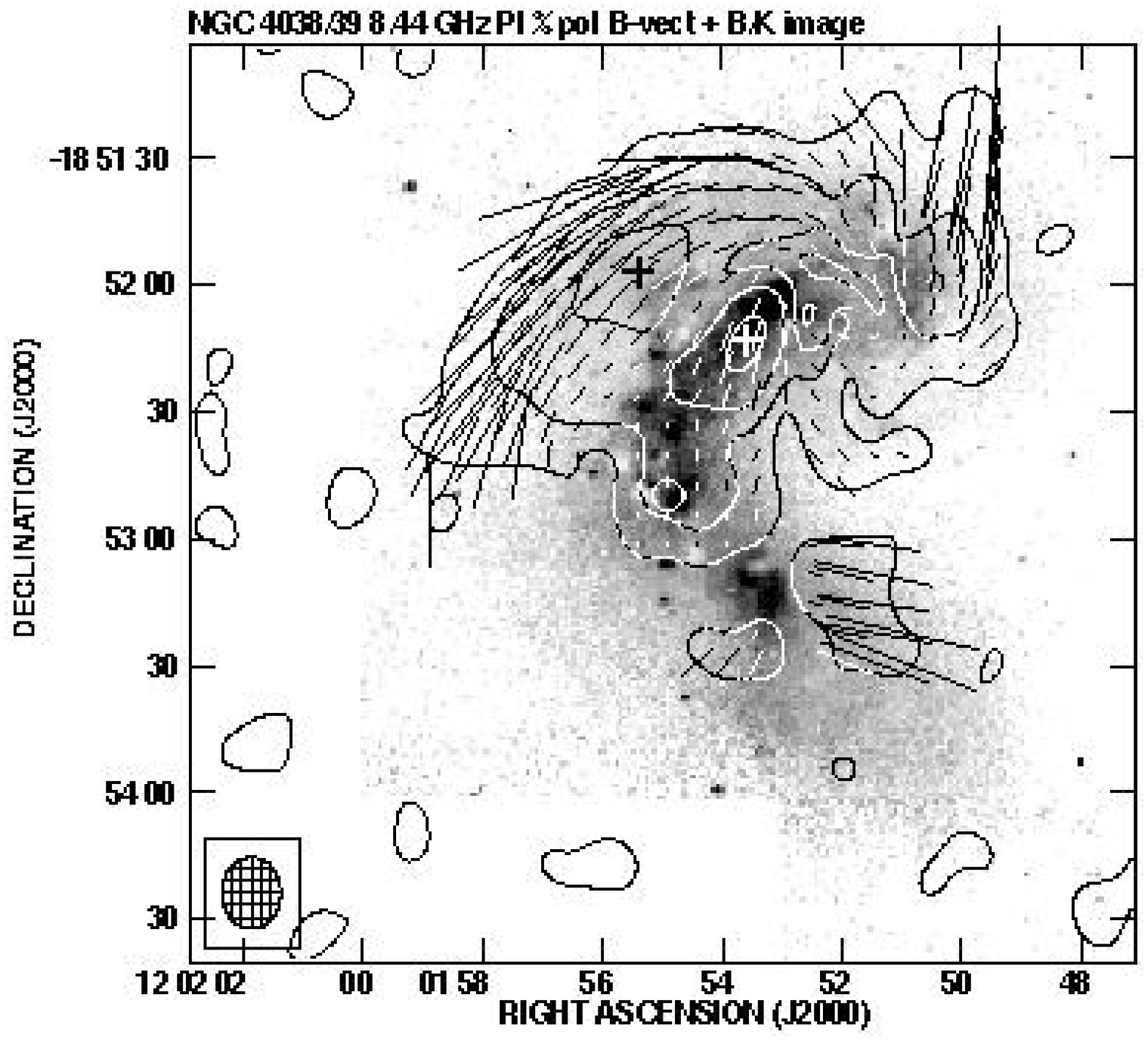}
\includegraphics[totalheight=7.9cm]{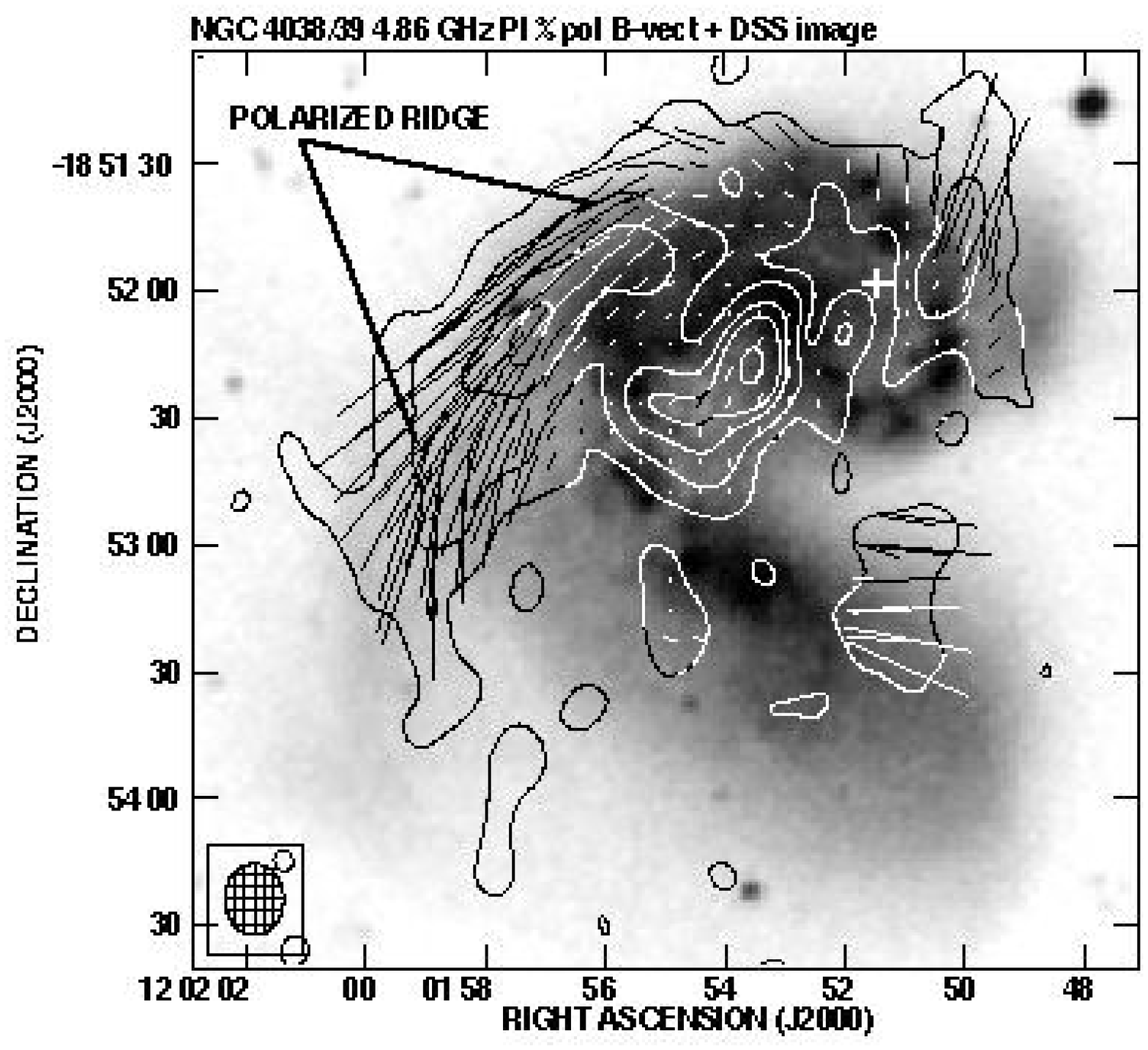}
\caption{
Left panel: The contours of polarized intensity with apparent B-vectors of 
polarization degree of NGC~4038/39 at 8.44~GHz overlaid upon the 
B/K colour index map (where dark areas indicates blue colour)
from Hibbard et al. (\cite{hibbard}). The contour levels are 
0.04, 0.13, 0.26, 0.46 mJy/b.a. and the rms noise is 0.013 mJy/b.a.
polarization degree of 12.5\%. The beam size is $17\arcsec\times 14 
\arcsec$. Right panel: contours of polarized intensity of NGC~4038/39 and 
apparent B-vectors of polarization degree at 4.86~GHz overlaid upon the DSS 
image. The contour levels are 0.04, 0.11, 0.21, 0.28, 0.42 mJy/b.a and the
rms noise is 0.014 mJy/b.a.
A vector of $10\arcsec$ corresponds to a polarization degree of 12.5\%. The 
beam size is $17\arcsec\times 14\arcsec$.
}
\label{pimap}
\end{minipage}
\end{figure*}
%=============================
%\clearpage

\subsection{Polarization maps}
\label{polmap}

The maps of polarized intensity at 8.44~GHz and 4.86~GHz (Fig.~\ref{tppi}) 
are suggestive for several structural components. In the NW disk (``inverted 9'') 
the polarized emission coincides with the optical one (Fig.~\ref{tppi} right 
panel). However, the orientations of apparent B-vectors 
do not follow the ``inverted 9'' feature. Instead, they form a rudimentary 
spiral shifted westwards from the northern galaxy nucleus,  being centered on 
RA$=12^\mathrm{h} 1^\mathrm{m} 51\fs6$ DEC$=-18\degr 51\arcmin 57\arcsec$ (Fig.~\ref{tppi} 
right panel, a white cross). The B-vectors at both frequencies are inclined by 
some $45\degr$ with respect to the total intensity ridge, thus, this feature is 
not caused by Faraday rotation effects. Close to the centre of northern galaxy 
the B-vectors are almost radial, following local dust lanes visible on HST 
image in Fig.~\ref{tphig}. Some traces of another spiral with a very large 
pitch angle are present around the nucleus of the southern galaxy. They are 
best visible at 8.44~GHz (Fig.~\ref{tppi} left panel). 

In the NE region of the interacting system the apparent B-vectors run parallel 
to a rather sharp, linear NE boundary of the optical emission. The map of 
polarized intensity at 8.44~GHz, much less affected by Faraday effects, shows 
a bright peak at RA$=12^\mathrm{h} 1^\mathrm{m} 55 \fs5$  DEC$=-18\degr 
51\arcmin 57\arcsec$ (Fig.~\ref{pimap} left panel, black cross). The 
polarization degree increases along this feature 
from about 5\% in its northern part to about 40\% at its southernmost tip. The 
polarized intensity at 4.86~GHz shows more clearly the southern part of this 
structure and forms a bright ridge extending along the NE boundary of the 
optical image of the system (Fig.~\ref{pimap} right panel).  

Another polarization component apparently extends along the complex of 
dark clouds visible in the B/K colour index image (Hibbard et al. 
\cite{hibbard}) shown in Fig.~\ref{pimap} (left panel). It is 
separated from the polarized ridge discussed above by a shallow 
depression, weakly visible at 8.44~GHz, but becoming much more 
conspicuous at 4.86~GHz. The northern part of the dark dust complex, 
devoid of observable H$\alpha$ emission (Whitmore et al. \cite{whitmore}) 
coincides with a local maximum of polarized intensity (see Fig.~\ref{pimap} 
left panel, white cross). However, at this position the polarization 
degree is rather moderate: about 10\% at 8.44~GHz and only 6\% at 4.85~GHz. 
The polarized emission at 8.44~GHz extends southwards along the region 
of strong reddening towards the bright total intensity peak (marked 
in the figure by a white circle). The polarization degree decreases 
southwards, becoming only 1\% at 8.44~GHz at the position of  the total 
intensity maximum. The polarized intensity at 4.86~GHz decreases southwards 
even faster. 

%=============================
\begin{figure} [t]
\resizebox{\hsize}{!}{\includegraphics{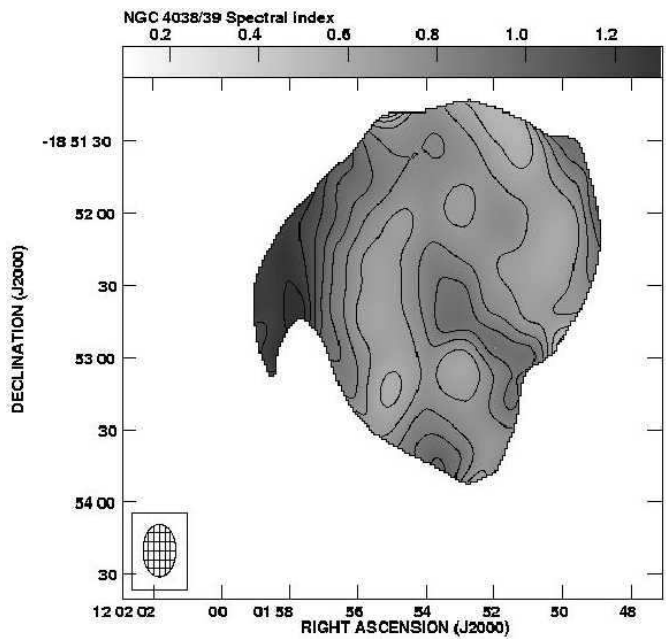}}
\caption{
The distribution of spectral index obtained by fitting a power low to 
the values of total intensity at 8.44, 4.86 and 1.49~GHz in NGC~4038/39. 
The maps of total intensity were convolved to a common beam of 
$22\arcsec\times 14\arcsec$. The contours are 0.4, 0.5, 0.6, 0.7, 0.8, 
0.9, 1.0, 1.1.} 
\label{spix}
\end{figure}
%=============================

\subsection{Spectral index distribution}
\label{spectral}

The spectral index distribution in NGC~4038/39 (Fig.~\ref{spix}) has been 
obtained by fitting for each map pixel a power law 
S$_{\nu}\propto\nu^{-\alpha}$ to the values of total intensity at 
three frequencies. The discussed interacting pair is characterized by 
strong variations of the spectral index over its surface. The western part 
of the ``inverted 9'' has a rather flat spectrum with $\alpha\simeq 0.6$, 
flattening northwards to 0.53. The radio spectrum also gradually flattens 
along the dark cloud complex in the central region, changing from about 
0.67 at its northern end to 0.58 in the radio bright region coincident with 
the southern  part of dust complex. Nuclear regions of both galaxies also 
have a rather flat spectrum with $\alpha$ of about 0.64. 

In contrast to that the gap between the galaxies, showing also depression 
in total radio intensity (Fig.~\ref{tppi}) has a spectral index of about 
0.94. A very steep spectrum has been found along the northeastern polarized 
region: most of it has $\alpha>1$, steepening towards the tidal tail to 
values of 1.2--1.3. There is also some spectral steepening across the ridge: 
from $\alpha\simeq$ 0.90--0.93 at its interface with a central cloud complex 
to values above 1.0 at the outer boundary. 

\subsection{Distribution of Faraday rotation and depolarization}
\label{faraday}

\begin{figure} [t]
\resizebox{\hsize}{!}{\includegraphics{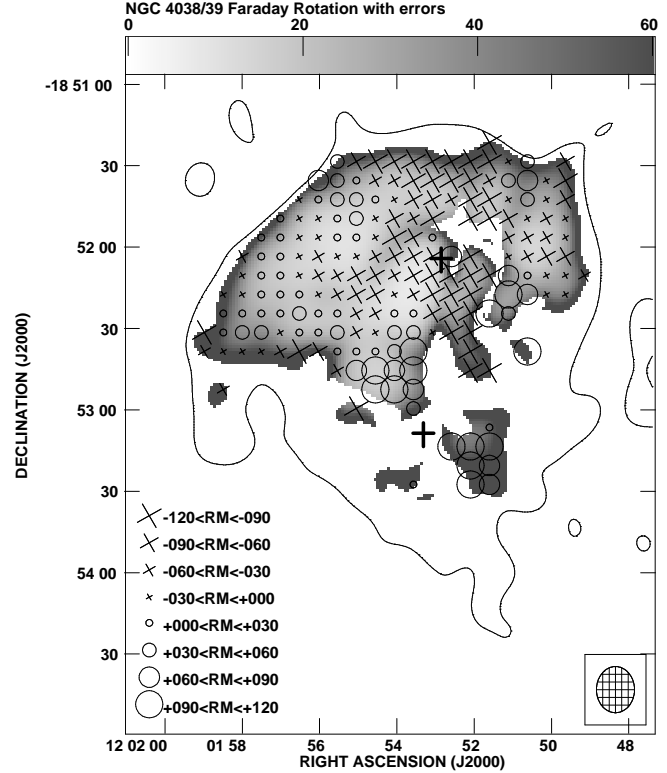}}
\caption{
Faraday rotation measure distribution in NGC~4038/39 computed between 8.44~GHz 
and 4.86~GHz plotted as symbols (see the Figure legend) overlaid upon the 
grayscale plot of rotation measure errors. The map resolution is 
$17\arcsec\times 14\arcsec$. Black crosses mark the positions of the
galaxy cores.
} 
\label{rm}
\end{figure}

\begin{figure} [t]
\resizebox{\hsize}{!}{\includegraphics{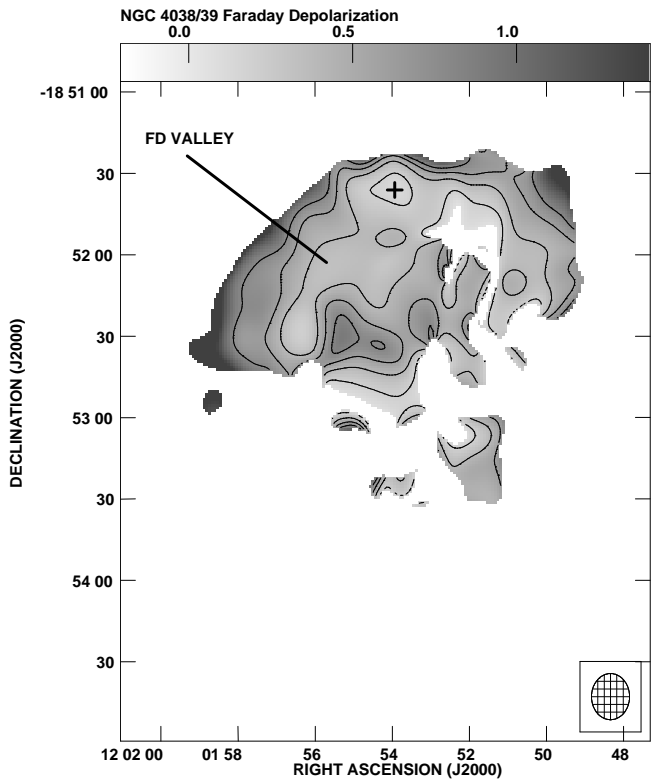}}
\caption{
Faraday depolarization distribution in NGC~4038/39 computed between 8.44~GHz 
and 4.86~GHz. The contours are 0.2, 0.4, 0.6, 0.8. The map resolution is 
$17\arcsec\times 14\arcsec$.} 
\label{fd}
\end{figure}

Maps of Faraday rotation and depolarization in NGC~4038/39 
(Figs.~\ref{rm} and \ref{fd}) have been 
computed between the frequencies of 4.86 and 8.44~GHz using the naturally 
weighted maps at the resolution of $14\arcsec\times 17\arcsec$. The galaxies 
lie at $287\degr$ Galactic longitude and $42.4\degr$ Galactic latitude where 
Simard-Normandin \& Kronberg (\cite{simard}) show low foreground rotation 
measures changing sign from place to place and absolute values 
$<30$\,rad\,m$^{-2}$. For this reason no correction for the foreground Faraday 
rotation has been applied. 

Over most of the eastern part of NGC~4038/39 and in the westernmost part of 
``inverted 9'' the Faraday rotation measures (RM) are generally small. Their 
absolute values in these regions exceed 50\,rad\,m$^2$ only in small areas 
comparable to the beam size (Fig.~\ref{rm}). The dividing line between the 
positive and negative values may depend on possible bias due to the foreground 
Faraday rotation. However, some statements do not depend on the uncertainty 
of foreground RM by some $\pm 30$\,rad\,m$^{-2}$. We note first the vertical 
belt of large negative RM's (between $-50$ and $-150$\,rad\,m$^{-2}$), about 
$30\arcsec$ wide, running north-south through the nucleus of the northern 
galaxy. Further to the south, around the southern galaxy nucleus, sudden RM 
jumps to strongly positive values up to $\ge+100$\,rad\,m$^{-2}$ are visible. 
We also note a substantial coherence of the sign of RM values. 
Independently of the assumed foreground rotation we get invariably a large 
domains of several beam areas having a constant sign of Faraday rotation 
measure and hence of the magnetic field direction.

%=============================
%=============================
%=============================
%=============================

In the eastern region of NGC~4038/39 the Faraday depolarization 
(FD=polarized intensity at~4.86GHz $/$ polarized intensity at~8.44GHz; Fig.~\ref{fd}) 
forms a valley with $\mathrm{FD}\simeq$0.2--0.3 along the channel occupied by the hot 
gas detected there by {\em Chandra} (Fabbiano et al. \cite{fabbiano}). Very strongly 
depolarized regions (FD$<0.2$) also appear near the nuclei of both galaxies. 
Another strong Faraday depolarization (FD$\simeq0.12$) occurs around 
RA$=12^\mathrm{h} 01^\mathrm{m} 54\fs 0$  DEC$=-18\degr 51\arcmin 35\arcsec$ (marked 
by a black cross in Fig.~\ref{fd}) where the Faraday rotation measure 
changes locally from $-120$\,rad\,m$^{-2}$ to +50\,rad\,m$^{-2}$.

\section{Discussion}
\label{discussion}
%==========================================
\begin{figure}[t]
\resizebox{\hsize}{!}{\includegraphics{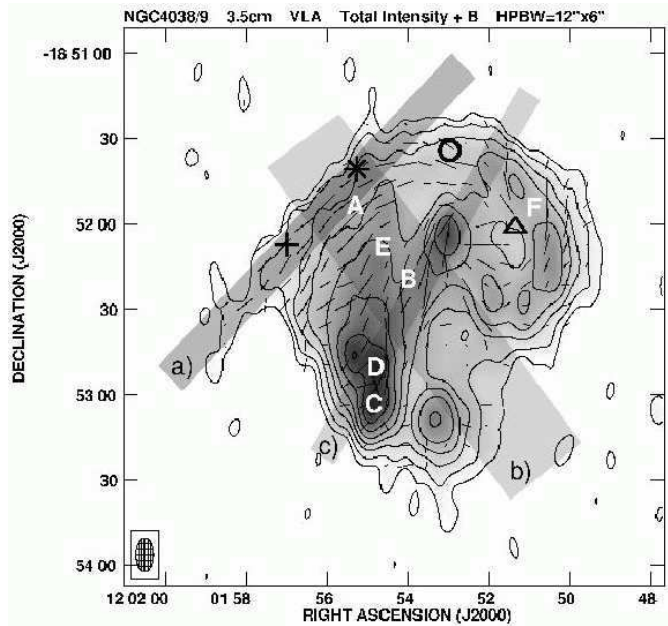}}
\caption{
Total radio intensity map and apparent B-vectors of polarized intensity of 
NGC~4038/39 at 4.86~GHz made from a combination of C and D array data, 
showing the location of particular radio emission regions (A--F) as discussed in 
Sect.~\ref{discussion}. The map resolution is $12 \arcsec \times 6\arcsec$. Shadow 
areas denote slices along which the profiles of various species were obtained 
(see Sect.~\ref{discussion}): a) a slice along northeastern ridge; b) a slice across 
the ridge; c) a slice along the ``overlapping region''. 
}
\label{sum}
\end{figure}

%=============================
\begin{figure} [ht]
\resizebox{\hsize}{!}{\includegraphics{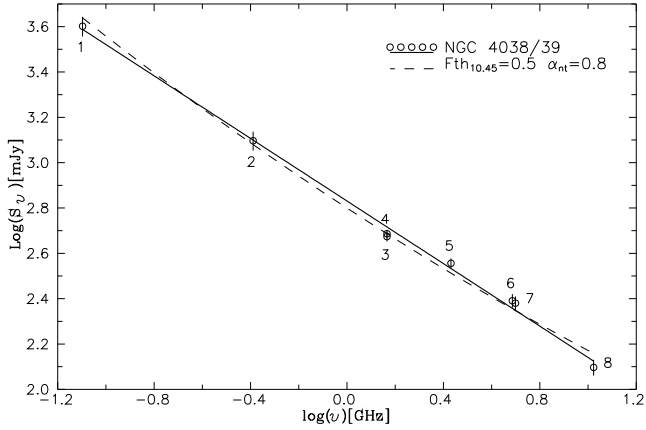}}
\caption{
The integrated spectrum of NGC~4038/39 compiled from the literature data: 1 -- 
Wright \& Otrupcek (\cite{wright}), 2 -- Wright \& Otrupcek (\cite{wright}), 
3 -- Hummel \& van der Hulst (\cite{hummel}), 4 -- Condon (\cite{condon83}), 
5 -- Wright \& Otrupcek (\cite{wright}), 6 -- Griffith et al.  
(\cite{griffith}), 7 -- Wright \& Otrupcek (\cite{wright}), 8 -- this paper. 
The dashed line shows the model spectrum assuming a thermal fraction at 
10.45~GHz of 50\% and a nonthermal spectral index of 0.8.
}
\label{spec}
\end{figure}
%=============================
\begin{figure} [t]
\resizebox{\hsize}{!}{\includegraphics{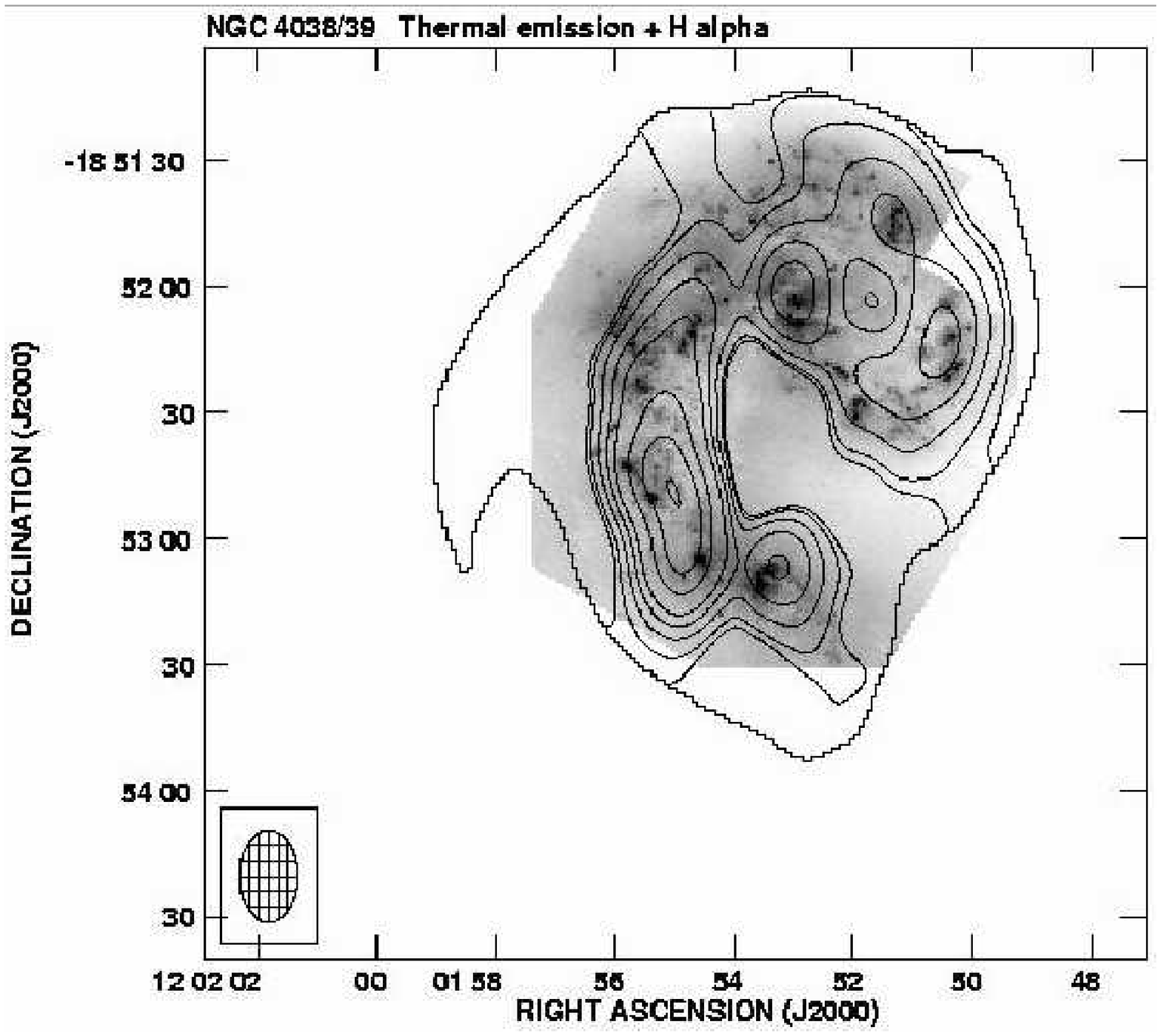}}
\caption{
The contour map of the thermal intensity at 8.44~GHz in NGC~4038/39 overlaid upon the 
grayscale image in the H$\alpha$ line made by HST (courtesy B.C. Whitmore from STScI). 
The contour levels are 0.1, 0.2, 0.7, 1.3, 2.2, 6.6, 8.8 mJy/b.a. and the map
resolution is $22\arcsec\times 14\arcsec$.
}
\label{therm}
\end{figure}
%================================================
\begin{figure} [ht]
\begin{center}
\resizebox{\hsize}{!}{\includegraphics{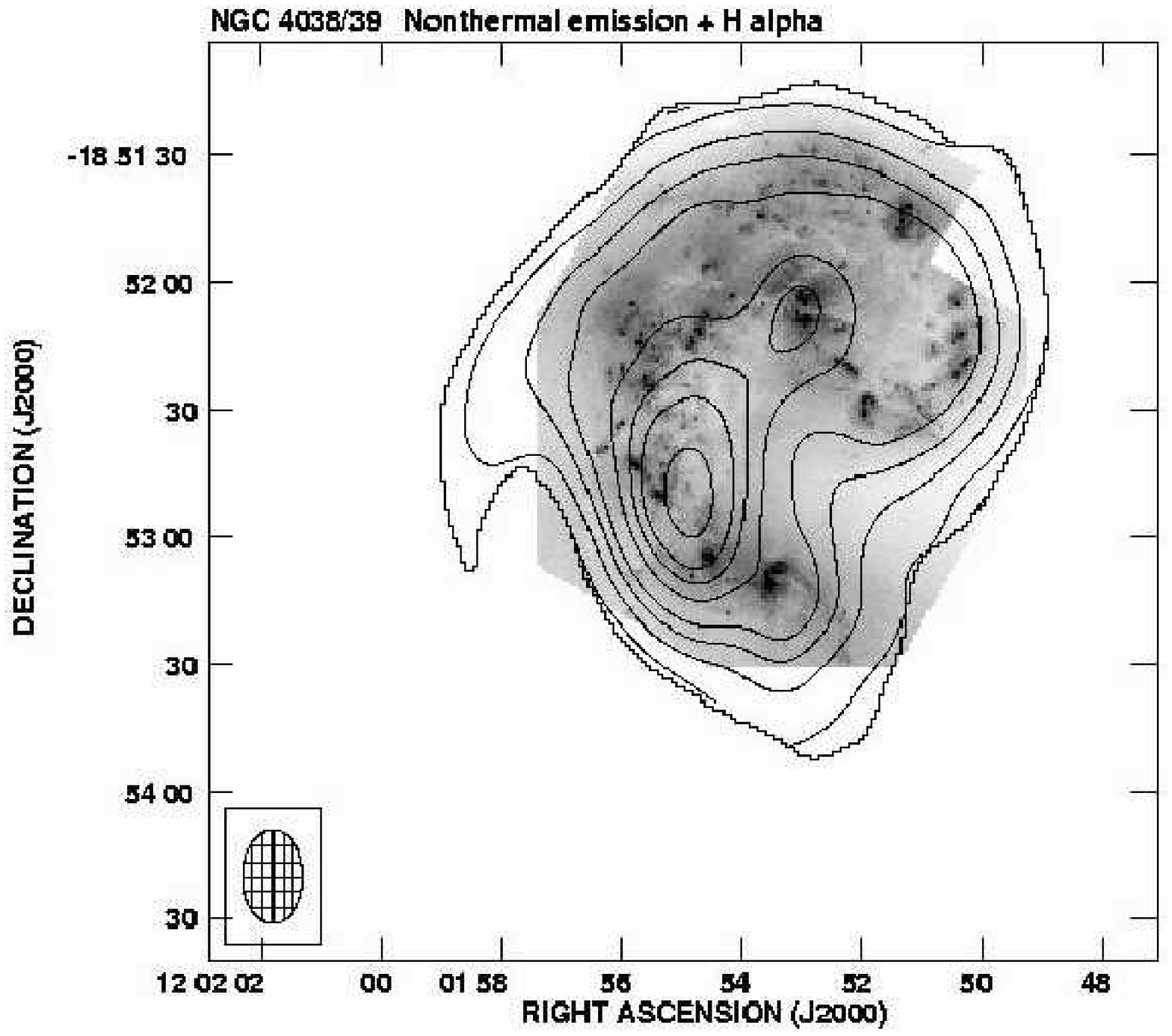}}
\end{center}
\caption{
The contour map of nonthermal intensity at 8.44~GHz from NGC~4038/39 overlaid 
upon the  grayscale image in the H$\alpha$ line made by HST (courtesy 
B.C. Whitmore from STScI). The contour levels are 0.1, 0.2, 0.7, 1.3, 2.2, 6.6, 
8.8 mJy/b.a. and the map resolution is $22\arcsec\times 14\arcsec$.
}
\label{notherm}
\end{figure}
%=============================
%=================================================
\begin{figure} [t]
\resizebox{\hsize}{!}{\includegraphics{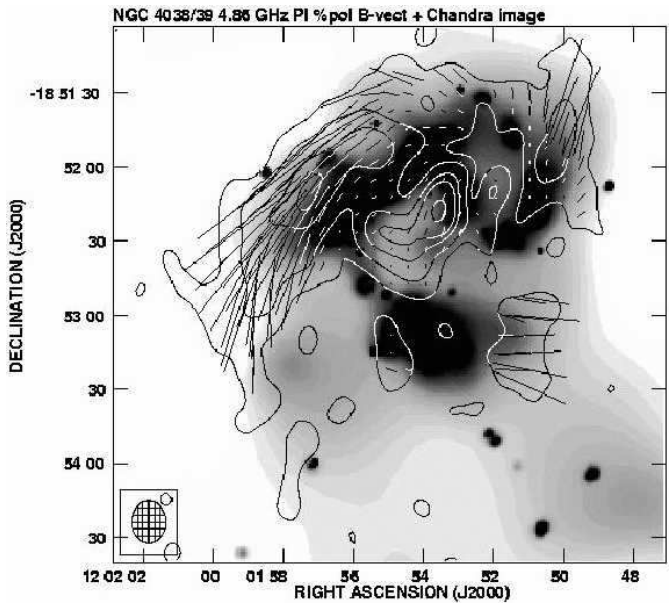}}
\caption{
Contour map of polarized intensity and apparent B-vectors of polarization
degree at 4.86~GHz overlaid upon the grayscale image of NGC~4038/39 in 
X-rays observed by {\em Chandra} satellite (Fabbiano et al. \cite{fabbiano}). 
The contour levels are 0.04, 0.11, 0.21, 0.28, 0.42 mJy/b.a. 
and the map resolution is $17 \arcsec \times 14\arcsec$.
}
\label{pixray}
\end{figure}

%===========================
\begin{figure} [ht]
\resizebox{\hsize}{!}{\includegraphics{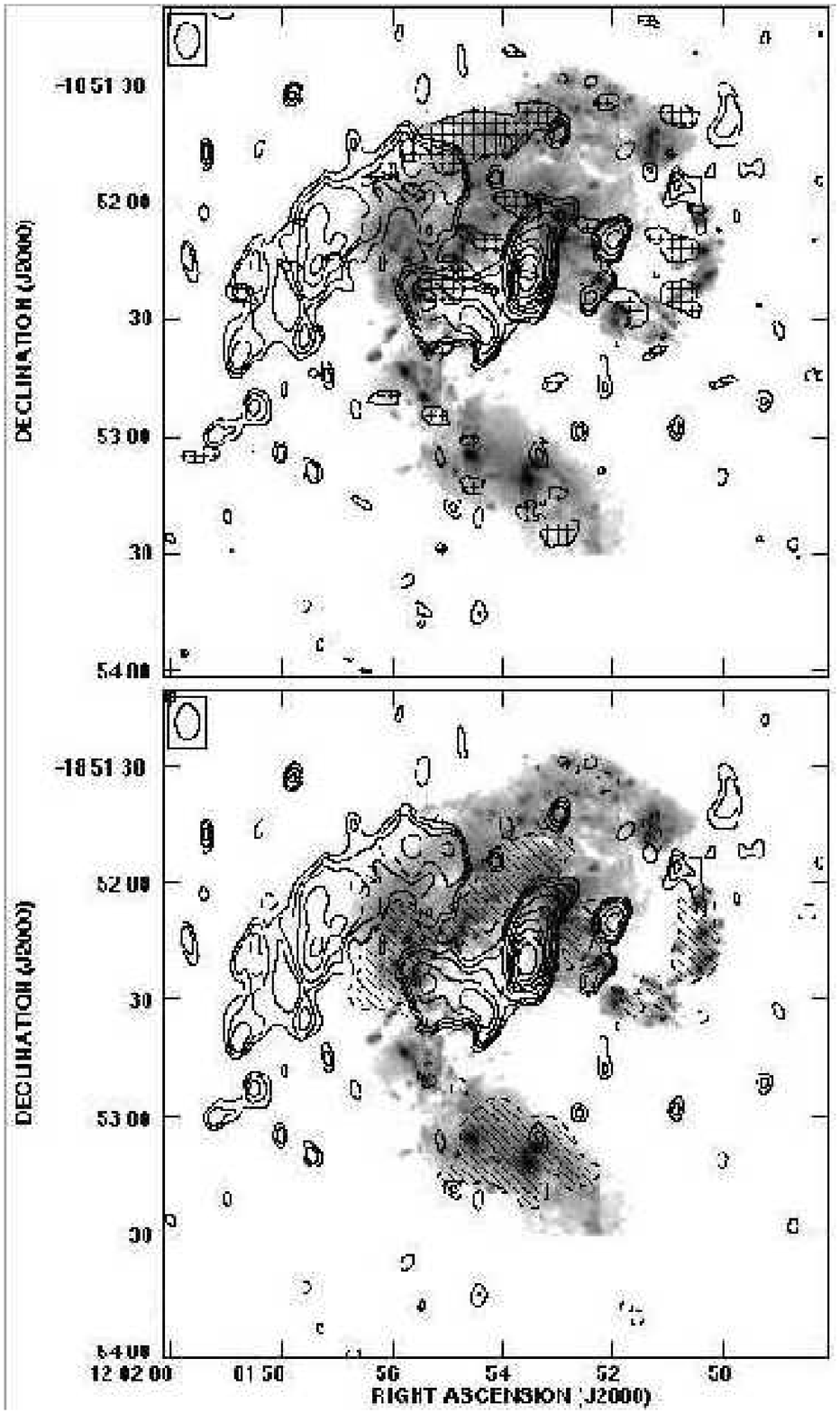}}
\caption{
Upper panel: contours of polarized intensity at 4.86~GHz in VLA C-array with 
uniform weighting (showing the brightest polarized structures at highest 
resolution) overlaid upon to the distribution of H$\alpha$ emission 
(grayscale). The cross-hatched area shows the region where HI emission was 
observed by Hibbard et al. (\cite{hibbard}) at the resolution of $11 
\arcsec\times 7 \farcs 4$ exceeds 100~mJy/b.a. Lower panel: the contours of 
polarized intensity and grayscale of H$\alpha$ emission (as above) and the 
region occupied by the diffuse X-ray emission (hatched area, Fabbiano et al. 
\cite{fabbiano}).
}
\label{combi}
\end{figure}
%================================================================================

To study the physical conditions and evolutionary stages of various locations 
in the Antennae system and interrelations of magnetic fields with particular 
phases of the interstellar medium, a comparison of our radio data with other 
spectral domains is necessary. For this purpose, mainly based on our radio 
maps in total and polarized intensity, we define several distinct regions 
labelled by letters A--F in Fig.~\ref{sum}. Region A is located in the middle 
of the polarized northeastern ridge, at the point where a bright ridge of 
polarized intensity coexists with a similarly oriented feature in \ion{H}{i} 
(Fig.~\ref{combi}). Regions B, C and D denote various parts of the so-called 
``overlap region'', constituting a dark cloud complex extending between the 
two interacting galaxies. Region B lies at its northern end while regions 
C and D denote the radio total intensity peaks at its southern tip 
(Fig.~\ref{tphig}). As region E we define the middle of the extended pool 
of X-ray emitting gas (Fig.~\ref{pixray}, Fig.~\ref{combi}). As region F 
we term the NW part of the relic spiral arm in the NW region of the system. 

In the following subsection, using our total intensity maps at 8.44, 4.86 and 
1.49~GHz, we first construct and present maps of thermal and nonthermal 
intensity of NGC~4038/39. We then analyze in next subsections the local 
interrelations of different ISM phases in regions A--F by comparing various 
distributions of radiation of different species along and across these 
particular regions. 

\subsection{Thermal and nonthermal intensity in NGC~4038/39}
\label{thnth}

Our total intensity maps at 8.44, 4.86 and 1.49~GHz allow to separate the
thermal and nonthermal radio emission components, assuming a constant nonthermal spectral 
index $\alpha_\mathrm{nt}$ over the whole radiation area. However, there are local
exceptions from this assumption where spectral index become particularly 
steep. There are regions in which the observed spectral index reaches the 
value of 1.0 (the radio ``valley'' between the galaxies on the western side 
of the system) or even 1.2--1.3 (the base of the tidal tail, Fig.~\ref{spix}), 
probably because of strong energy losses by CR electrons. 

%======================================
\begin{figure}[t]
\resizebox{\hsize}{!}{\includegraphics{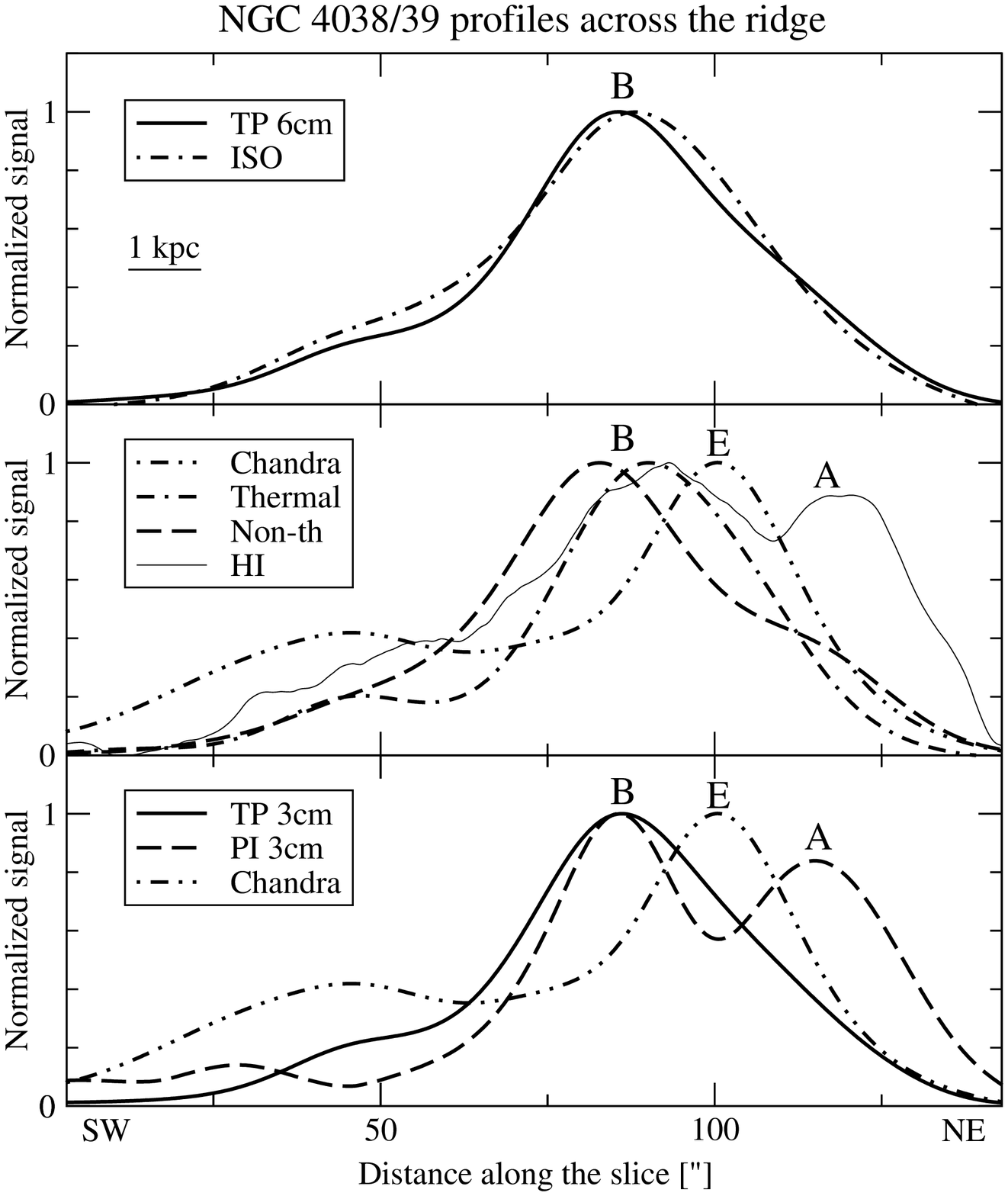}}
\caption{
The profiles of various species across the NE nonthermal ridge integrated 
along the slice b) marked in Fig.~\ref{sum}.
}
\label{across}
\end{figure}

To compute the maps of thermal and nonthermal intensity (and thus also the map 
of thermal fraction) we tried to find the best value which could be adopted 
for mean $\alpha_\mathrm{nt}$ in possibly largest area of the disks ignoring 
mentioned regions of strong electron energy losses. For this 
purpose we used the integrated spectrum of NGC~4038/39 
(Fig.~\ref{spec}) constructed from the available literature data and  
our own measurements. The flux density at 10.45~GHz was obtained by 
integrating our Effelsberg map in circular rings. The spectrum can be equally 
well represented by a whole range of combinations of nonthermal spectral 
index and mean thermal fraction at 10.45~GHz ranging from $\alpha_\mathrm{nt}$ = 0.69 
$\overline{f_\mathrm{th}}$ = 0 to $\alpha_\mathrm{nt}$ = 0.8 and $\overline{f_\mathrm{th}}$ = 0.5. 
All fits between these values lie well within 3$\sigma$ flux density errors 
(Fig.~\ref{spec}) and yield very similar r.m.s. residuals. A further increase 
of a mean thermal fraction (hence a steeper nonthermal spectrum) causes a 
rapid increase of residuals and yields an unrealistically curved total spectrum. 
On the other hand, adopting a flatter nonthermal spectrum ($\alpha_\mathrm{nt}<0.8$) 
seems unreasonable in computing the maps because large disks areas are occupied 
by a spectrum steeper than this value. We used thus a compromise assumption of 
$\alpha_\mathrm{nt}$=0.8 and computed maps of thermal and nonthermal intensity as well as
the map of thermal fraction at 8.44~GHz, excluding from the analysis regions 
with a steeper spectrum. While this assumption is valid for the area covered 
by optically bright galaxies, the nonthermal spectrum is definitely steeper in 
the vicinity of the radio extension along the northeastern tail and in the radio 
``valley'' between the galaxies. 

%=======================================
\begin{figure}[t]
\resizebox{\hsize}{!}{\includegraphics{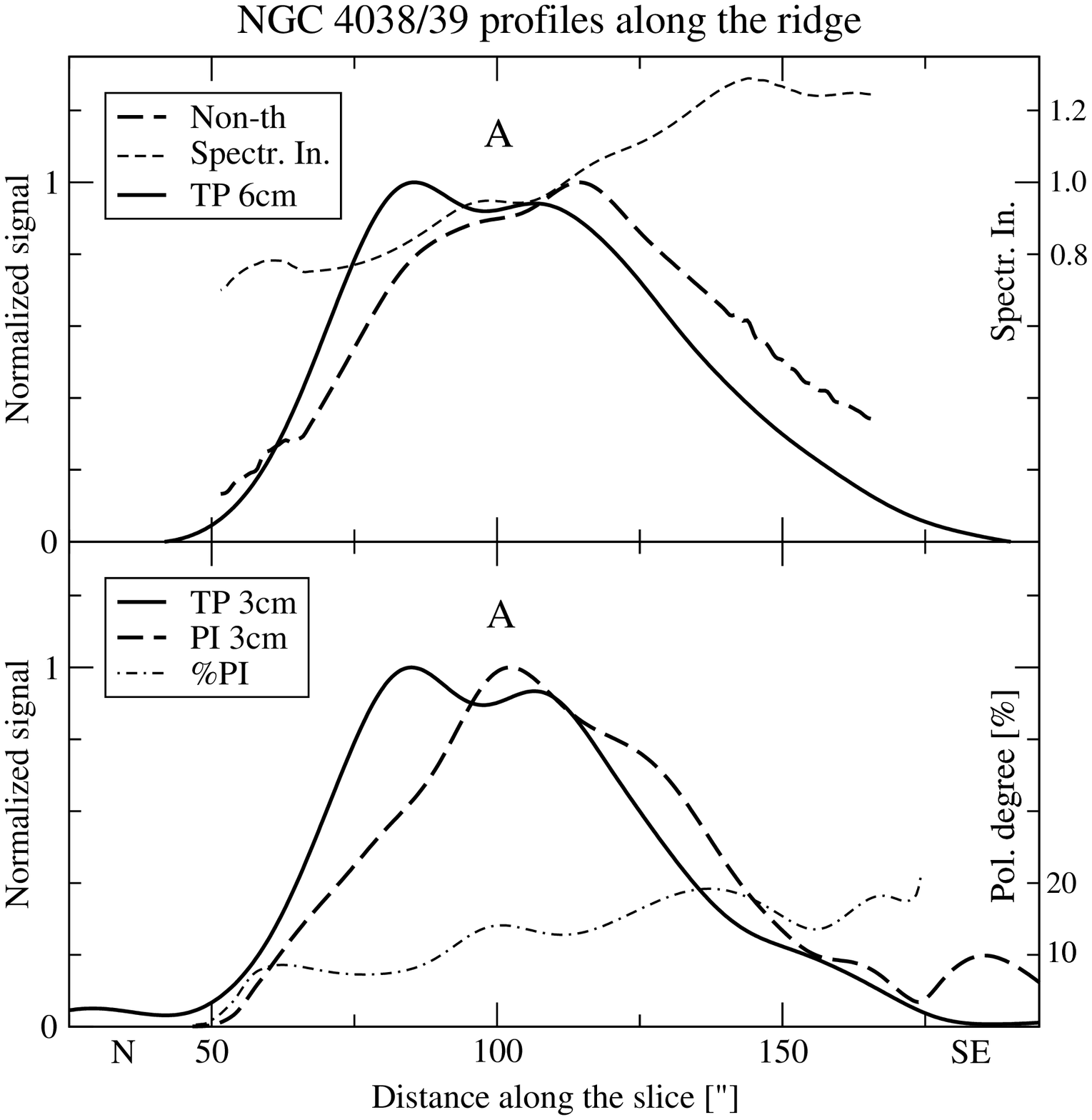}}
\caption{
The profiles of various species along the NE nonthermal ridge integrated 
along the slice a) marked in Fig.~\ref{sum}.
}
\label{along}
\end{figure}
%=========================================

We note here that a mean thermal fraction of 0.5 at 10.45~GHz implied by this 
assumption is larger than the mean value of $0.30\pm 0.05$ at 10~GHz estimated 
for 74 galaxies from the Shapley-Ames survey by Niklas et al. (\cite{niklas}). 
This is not surprising as the pair NGC~4038/39 actively forms stars and produces 
ionized gas very efficiently.

Regions of substantial thermal fraction (f$_{th}$ at 8.44~GHz of 30\%--40\%) 
are found at the positions of the nuclei, in the western part of the 
``inverted 9'' (region F), and along the dark cloud complex. In the southern 
part of the obscured regions  C and D the thermal fraction reaches  some 
47\%.  The distribution of thermal brightness at 8.44~GHz is compared to the 
distribution of \ion{H}{ii} regions in Fig.~\ref{therm}. The ridge of enhanced 
thermal intensity follows well the chain of \ion{H}{ii} regions in the 
``inverted 9'', making some minimum  east of the northern galaxy nucleus,  
(close to our region B) then following the  belt of \ion{H}{ii} regions 
along the central dark cloud. The thermal brightness has a maximum in the 
southern part of the cloud, while no particular peak in H$\alpha$ emission 
nor any concentration of \ion{H}{ii} regions is seen. This is discussed 
further in Sect.~\ref{cloud}.

The nonthermal intensity distribution (Fig.~\ref{notherm}) is smoother and fills 
the whole system, as expected from smearing small-scale features by the 
cosmic ray propagation. The southern part of the central dark cloud shows also 
a substantial enhancement of nonthermal intensity. 

\subsubsection{Star formation in the vicinity of the northeastern ridge}
\label{ridge}

The profile of the total and polarized radio intensity compared to that of 
\ion{H}{i}, X-ray and the radio thermal and nonthermal intensity {\em across}
the northeastern ridge is presented in Fig.~\ref{across}. Most of species  
peak at the distance from  the slice origin $x \simeq 80\arcsec$ (region B in 
Fig.~\ref{sum}). This corresponds to the northern part of the dark cloud 
complex extending between the galaxies, flanked on its NE side by a 
star-forming region. The separation of thermal and nonthermal intensity 
is visible in the middle panel of Fig.~\ref{across}. The high resolution 
optical image (shown in Fig.~\ref{tphig}) helped us to associate the thermal 
emission peak with the star-forming clump and the nonthermal one with a dusty 
cloud. This means that there is not much hidden star formation in the cloud 
itself. Instead there is a strong, partly regular magnetic field (note the 
peak of polarized intensity) into which the electrons produced in the 
star-forming region propagate, yielding enhanced nonthermal intensity. 
We note that the distribution of the $\lambda15\mu$m brightness (obtained
by Mirabel et al. \cite{mirabel} with ISO) agrees with 
the total (i.e. sum of thermal and nonthermal) intensity, its peak being 
located between the maxima of these two emission components. 
Both thermal and nonthermal intensity decrease towards the NE. 

The X-ray emission peaks about $x=100\arcsec$ in a pool of hot gas 
(region E). This ``X-ray
ridge''  does not correspond to any increase in total radio intensity.
According to  Zhang et al. (\cite{zhang}) this region is filled mostly
with  intermediate age ($3<t<16~$Myr) stellar clusters. It may be a
relic of an old star formation burst where the cosmic ray
electrons lost most of their energy, making no significant contribution
to the nonthermal intensity. Synchrotron lifetime of electrons with
$\simeq 4$~GeV energy in a 23$\,\mu$G field (Table~\ref{comp}) is 
$\simeq 5\times 10^6$~yr which gives a lower limit for the age of the 
starburst. 

Polarized intensity in region E shows a striking minimum at 8.44~GHz
and at 4.86~GHz (see Fig.~\ref{pimap}) together with strong 
wavelength-dependent depolarization (FD ``valley'' in Fig.~\ref{fd}). 
At 4.86~GHz the degree of polarization is reduced by a factor of 0.35
relative to 8.44~GHz. As Faraday rotation there is smaller than 
50~rad $\mathrm{m}^{-2}$ (Fig.~5), depolarization due to differential
Faraday rotation cannot be more than by a factor of 0.98. Thus, the 
wavelength-dependent depolarization is due to Faraday dispersion (Burn \cite{burn},
Sokoloff et al. \cite{sokoloff}). A RM dispersion of $\sigma_\mathrm{RM}^2$
of $1.1 \times 10^5$ yields a depolarization of 0.30 at 4.86~GHz and 0.84
at 8.44~GHz giving the observed ratio of 0.35, but insufficient to 
explain the minimum in polarized intensity in region E at 8.44~GHz.
In that case additional wavelength-independent beam depolarization by
an increased degree of small-scale magnetic field tangling is required.

\subsubsection{Magnetic fields and cosmic rays in the northeastern ridge}
\label{magneticridge}

At $x\simeq 120\arcsec$ (region A) a peak of polarized intensity is 
observed, coinciding with a \ion{H}{i} maximum (Fig.~\ref{across}, see also 
Fig.~\ref{combi}). It does not correspond to any rise of other constituents. 
According to recent numerical simulations kindly supplied to us by Dr 
Englmaier (priv. comm., see also Englmaier et al. \cite{englmaier}) this 
region constitutes the beginning of the tidal (both stellar and gaseous) tail, 
unfolding towards the east and northeast. Because of little star formation 
this region does not correspond to any particular total intensity features, 
instead it is well revealed by the polarized intensity because of highly 
ordered magnetic fields. According to the suggestion by Fabbiano et 
 al. (\cite{fabbiano03}) this region may be subject to external compression by 
the ambient gas pushing inwards the pool of hot gas. This may imply that 
our region A, with well aligned polarization vectors in front of the X-ray 
maximum, could be the result of magnetic field compression. Gas infall in 
this region has also been suggested by Hibbard et al. (\cite{hibbard}). 
However, the high resolution map of polarized intensity (Fig.~\ref{combi}) 
does not show any narrow ridge, as expected in case of external compression. 
Moreover, the highly polarized region is significantly displaced 
southeastwards with respect to the X-ray ridge and also to the \ion{H}{i} 
ridge (Fig.~\ref{combi}). It is more likely that the increased degree of 
magnetic field regularity is caused by field geometry and decaying 
turbulence and/or  shearing gas motions (see below) than by gas infall 
pushing inwards the X-ray emitting matter.

Fig.~\ref{along} shows the distribution of various constituents {\em along} 
the discussed polarized ridge. There is still some weak star formation in its 
N part, as indicated by a total intensity peak at $x \simeq 80\arcsec$. 
The polarized intensity peaks more southwards at $\simeq 100\arcsec$ at 8.44~GHz 
(Fig.~\ref{along}) and $x \simeq 120\arcsec$ at 4.85~GHz. While both 
total and polarized intensity decrease towards the
tidal tail, the polarized intensity fades somewhat farther away than
the total intensity (Fig.~\ref{along}, lower panel). This means that the magnetic 
field gets more ordered (the polarization degree is increasing) as
the turbulent motions decay away from star-forming regions. 
A substantial part of a high polarization of this region may come from a 
``pseudo-regular field'', i.e. a highly anisotropic random one, stretched by 
shear accompanying the unfolding motion of the tidal tail. Such a field structure 
yields a high degree of polarization, not accompanied by strong Faraday rotation, 
because the stretching of random field alone cannot generate coherent 
uni-directional magnetic fields. We found indeed that Faraday rotation 
in the southern, weakly star-forming part of the polarized ridge is 
small ($\le 30$\,rad\,m$^{-2}$, see Fig.~\ref{rm}).

However, the Faraday rotation measures (in absolute values) increase above 
100\,rad\,m$^{-2}$ along the ridge of polarized emission towards the north. 
These strongly non-zero Faraday rotation indicate that we deal with 
genuinely regular magnetic fields. Together with the increase in RM, the 
degree of polarization decreases. This anticorrelation suggests 
that we deal with the same large-scale coherent magnetic field and 
variations in the orientations of magnetic lines are responsible for the 
observed variations of RM and the field regularity. This explains the high 
Faraday rotation in the north. The small degree of polarization (low 
field regularity) in the north is in line with the high star formation 
rate, a relation also observed in other regions of NGC~4038/39 (see 
Sect.~\ref{cloud}). We believe that the southern region also contains 
strong coherent fields but they run almost parallel to the sky giving
little Faraday rotation.

A simple model reproducing simultaneously Faraday rotation measures, 
polarized intensities and thermal intensity at 8.44~GHz has been 
constructed. We assume that 20\% of the observed thermal intensity 
(Fig.~\ref{therm}) is due to 
diffuse ionized gas responsible for Faraday rotation. Varying the pathlength 
of thermal emission between 100 and 400~pc, the nonthermal pathlength between 
1\,kpc and 2\,kpc, and the ionized gas filling factor between 0.05 and 0.1 
(see Ehle \& Beck \cite{ehle}) we found that all radio data in 
the SE part of the polarized ridge (marked by a black cross in Fig.~\ref{sum}) 
 require a regular field of about $10\pm1\,\mu$G (see Table~\ref{comp}), 
inclined to the sky plane by some $4\pm 2\degr$, 
only weakly depending on the assumptions listed above. This result suggests 
that most of the regular magnetic field lies in the sky plane. 

In the northern part of the ridge (marked by a black circle in 
Fig.~\ref{sum}),  showing higher RM, the same analysis implies that 
the regular field of about $9\pm 2\,\mu$G is running away from the observer 
at the angle of $59\pm 9\degr$ to the sky plane. From this place 
towards the SE (in a region marked by a black triangle in Fig.~\ref{sum}) 
the regular magnetic field changes its direction. It remains inclined 
to the sky by some $34\pm 9\degr$  but points towards us. Our model 
indicates that we generally deal with a three-dimensionally curved 
structure of magnetic fields.

These results suggests that we need a higher degree of coherence
in the southern part of the NE ridge as expected for decaying turbulent 
motions away from star-forming regions. While the above mechanism 
would be the main cause of coherent fields, the shearing motions may 
align the regular field along the tail.

Along the nonthermal ridge the radio spectrum (Fig.~\ref{along}) 
gradually steepens (also the nonthermal one as discussed in Sect.~\ref{thnth}), 
indicating an aging population of CR electrons as they propagate further away 
from their sources in star-forming regions. The proximity of a hidden strong 
star formation nest in the southern part of the ``overlapping region'' (D 
in Fig.~\ref{sum}) does not result in an increased CR electron content in the 
ridge, probably because of a strong regular magnetic field perpendicular to 
the required direction of CR propagation.

\subsubsection{The dark cloud complex}
\label{cloud}

\begin{figure}[t]
\resizebox{\hsize}{!}{\includegraphics{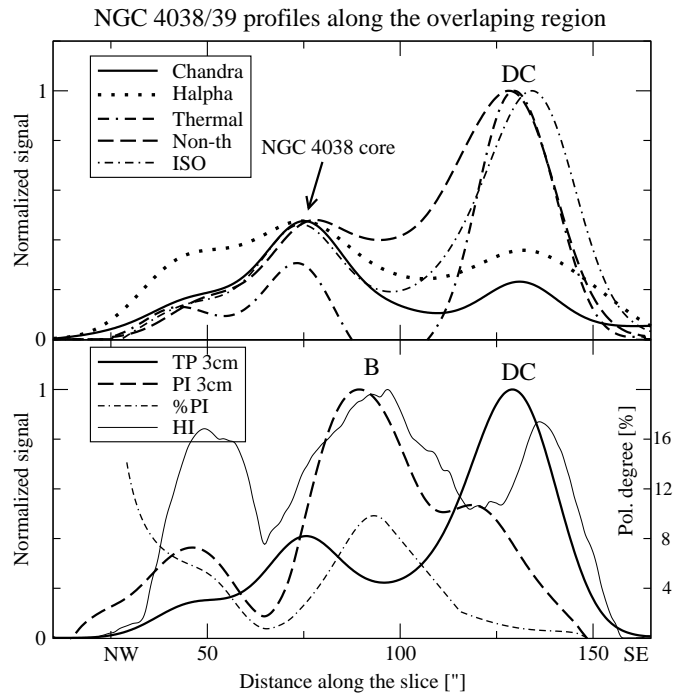}}
\caption{
The profiles of various species in the ``overlapping region''
integrated along the slice c) marked in Fig.~\ref{sum}
}
\label{overlap}
\end{figure}

Fig.~\ref{overlap} shows the brightness variations of various constituents 
along the dark cloud complex extending to SE from the nucleus of the northern 
galaxy (seen on e.g. the B/K colour index map in Fig.~\ref{pimap}, left panel).
Starting from the northwest there is a small peak in polarization 
($x \simeq$50\arcsec), corresponding to the interface between the rudimentary 
magnetic spiral in the NW disk (see Sect.~\ref{polmap}) and the polarized 
ridge discussed in Sect.~\ref{ridge}. The whole relic spiral associated 
with the ``inverted 9'' optical feature deserves attention, too. In this 
region the observed B-vectors have almost a constant inclination 
($\simeq 30\degr$--$40\degr$) with respect to the narrow blue optical ridge, 
thus the magnetic field cannot be simply aligned by compression as this
structure unfolds. We suspect this feature to be the relic of the galaxy's 
original magnetic field. However, Faraday rotation measure in the area of 
this rudimentary spiral undergoes strong variations and jumps, changing 
suddenly between $\simeq-100$\,rad\,m$^{-2}$ and $\simeq +100$\,rad\,m$^{-2}$. 
This indicates that the magnetic spiral has in fact a complex three-dimensional 
structure as does also the velocity field (Barnes \cite{barnes}).
Similar but weakly visible spiral with a 
large pitch angle is visible also around the southern nucleus.

Moving to the southeast, at $x\simeq 75\arcsec$ (Fig.~\ref{overlap}) there 
is a small peak in almost all species (except polarization and \ion{H}{i}) 
corresponding to an enhanced star formation around the nucleus of
the northern galaxy. The polarized intensity continues to rise
towards the SE, reaching a strong peak (as does the \ion{H}{i}) at
$x\simeq$ $85\arcsec$ (region B) while all other quantities, sensitive
to the star-forming activity show a minimum close to this place. This
region corresponds to the northern part of the dark cloud complex. Our
observations imply thus that there is no hidden strong star formation
there. In fact the sub-mm observations with SCUBA (Haas et al.
\cite{haas}) show that this region is radiating quite strongly at $\lambda
850\mu$m being very weak at $\lambda 450\mu$m, which indicates
its low dust temperature.

More to the south at $x\simeq130\arcsec$ the radio thermal and nonthermal
brightness, as well as the intensity at $\lambda 15\mu$m (ISO) show a strong peak
(Fig.~\ref{overlap}), constituting a beam-smeared 
blend of regions C and D as defined in Fig.~\ref{sum}. The peaks of the radio and 
infrared emission are more than twice stronger compared to those at the 
position of the nucleus of northern galaxy. In contrast, the H$\alpha$ and 
X-ray peaks are considerably weaker than at the northern nucleus. In case of 
H$\alpha$ emission this means strong extinction in the optical domain, hiding 
a very strong burst of star formation. This agrees with results by Haas et 
al. (\cite{haas}) who find there strong emission at $\lambda 450\mu$m 
indicating  considerably warmer dust. The peak of thermal radio 
intensity is particularly appealing. {\it The radio thermal intensity 
can therefore serve as a 
good tracer of thermal gas ionized by recent star formation, but still hidden
in the optical domain by heavy extinction in a dusty molecular environment}. 

Whitmore \& Zhang (\cite{whitmore}) argued that most radio emission 
from the radio-bright phase of young star clusters in the Antennae originate 
from H$\alpha$-emitting gas. Our results show that their conclusion is 
not valid for the extended radio emission. Indeed, the compact sources 
comprise only 11\% of the total radio intensity studied here. The 
distinct behavior of H$\alpha$ and total radio intensity is visible 
for example in Fig.~\ref{overlap} where their profiles along the 
``overlapping region'' show different shapes.

To explain a weak X-ray emission in the southern part of the dark cloud complex 
we have two possibilities: either strong extinction by a \ion{H}{i} concentration, 
or the star formation burst may be so young that did it not yet produce a large 
reservoir of diffuse hot gas but sufficiently old to produce radio-emitting CR 
electrons, hence its age is between $10^6$~yr and $10^7$~yr. The discrimination 
between these possibilities is impossible without analyzing the X-ray spectra,
which is beyond the scope of this paper.

%==================================================================================
\begin{table*}
\begin{center}
\caption{Magnetic fields strength and field regularity in different regions of
the merging system of NGC~4038/39.
}
\label{comp}
\begin{tabular}{lrrrr}
\hline\hline\noalign{\smallskip}
Region                               & inclination &  $B_{\mathrm{tot}}$   &   $B_{\mathrm{reg}}$  & regularity\\
                                     & ($\degr$)~~~~~ & ($\mu$G) & ($\mu$G) & $B_{\mathrm{reg}}/B_{\mathrm{ran}}$ \\
\hline
The whole merging system             & 64 & 20 & 5 & 0.25 \\
                                     &  0 & 24 & 5 & 0.19 \\
NGC~4038 (without overlapping region)  & 55 & 21 & 5 & 0.25 \\
NGC~4039 (without overlapping region) & 73 & 15 & 3 & 0.19 \\
Overlapping region as a whole        & 64 & 27 & 4 & 0.16 \\
                                     &  0 & 34 & 4 & 0.12 \\
Central peak of NGC~4038              & 55 & 29 & 8 & 0.30 \\
Central peak of NGC~4039             & 73 & 21 & 1 & 0.05 \\
N part of the northeastern ridge, modelled in Sect.~\ref{magneticridge}       & 59 & 19 &  9 & 0.52 \\
(RA$=12^{\mathrm{h}} 01^{\mathrm{m}} 53\fs 0$ Dec$=-18\degr 51\arcmin 
33 \arcsec$, a black circle in Fig.~\ref{sum})  & \\
Middle part of the northeastern ridge, modelled in Sect.~\ref{magneticridge}       & 34 & 18 &  8 & 0.48 \\
(RA$=12^{\mathrm{h}} 01^{\mathrm{m}} 55\fs 3$ Dec$=-18\degr 51\arcmin 
41 \arcsec$, a black star in Fig.~\ref{sum})  & \\
SE part of the northeastern ridge, modelled in Sect.~\ref{magneticridge}       &  4 & 16 & 10 & 0.88 \\
(RA$=12^{\mathrm{h}} 01^{\mathrm{m}} 57\fs 0$ Dec$=-18\degr 52\arcmin 
06 \arcsec$, a black cross in Fig.~\ref{sum})  & \\
Region A, middle of the northeastern ridge & 55 & 22 & 13 & 0.76 \\
                                     &  0 & 24 & 10 & 0.45 \\
Region B, overlapping region, north   & 55 & 27 & 12 & 0.50 \\
                                     &  0 & 30 & 8 & 0.29 \\
Region C, overlapping region, south   & 64 & 30 & 3 & 0.12 \\
                                     &  0 & 37 & 2 & 0.05 \\
Region D, overlapping region, south   & 64 & 32 & 7 & 0.23 \\
                                     &  0 & 39 & 4 & 0.10 \\
Region E, X-ray ridge                & 55 & 24 &  8 & 0.35 \\
                                     &  0 & 27 & 5 & 0.20 \\
Region F, spiral arm in NGC~4038     & 55 & 23 & 9 & 0.43 \\
``Interarm'' region in NGC~4038      & 55 & 22 & 5 & 0.21 \\
(RA$=12^{\mathrm{h}} 01^{\mathrm{m}} 51\fs 4$ Dec$=-18\degr 52\arcmin 
0 \arcsec$, a black triangle in Fig.~\ref{sum})  & \\
\hline
\end{tabular}
\end{center}
\end{table*}
%==================================================================================

The degree of polarization and the polarized intensity is highest in the 
cool northern  part of the ``overlapping region'' (close to B). They both
decrease gradually towards the SE while  the total (i.e. the sum of thermal 
and nonthermal) intensity continuously rises (Fig.~\ref{overlap}). This  
implies that in the southern region of the dark cloud complex the regular 
magnetic field becomes more efficiently destroyed by hidden star formation. 
We also note that the B-vectors keep a coherent structure along the whole 
region (Fig.~\ref{pimap}).  We deal here with genuinely  regular fields: the 
Faraday rotation measure keeps constantly high absolute values  ($\ge $
100\,rad\,m$^{-2}$) in two large coherent domains: negative in the northern part,  
changing to positive in its southern region (Fig.~\ref{rm}). The analysis 
done in the same way and under the same assumptions as in Sect.~\ref{magneticridge} 
shows that equipartition regular magnetic field in the northern part (region B)  
is about $B_{\mathrm{reg}}\simeq 5\,\mu$G (as estimated from the polarized 
brightness at 8.44~GHz) and is inclined to the  sky  
plane by some $37\pm5\degr$. This is implied by the Faraday rotation 
measure and thermal gas density derived from the thermal intensity.
The sudden change of RM in the middle of the cloud complex also means a sign reversal
of the line-of-sight regular field  component. We tentatively speculate
that this might be a meeting point of the global regular magnetic fields
of the two galaxies.

As also noted by Haas et al. (\cite{haas}) the southernmost tip of the 
dark complex (region C, distinguished in our high resolution total radio 
intensity map, Fig.~\ref{tphig}) has a several times higher ratio of 
$\lambda 15\mu$m-to-radio intensity than region D. This is also visible in 
Fig.~\ref{overlap} (upper panel) as a shift of 
the infrared peak with respect to the radio emission. It is possible that our 
region C contains a hidden, strongly star-forming nest residing in the 
environment of weaker total magnetic field or the star formation burst is 
so young ($<10^6$~yr) that not enough CR electrons have been produced so far.

\subsection{Magnetic field strengths}
\label{strength}

For the detailed analysis of magnetic fields in NGC~4038/39 we performed  
calculations of total $B_{\mathrm{tot}}$ and regular $B_{\mathrm{reg}}$
equipartition magnetic field strengths and of the degree of field regularity 
$B_{\mathrm{reg}}/B_{\mathrm{ran}}$ (the ratio of regular to random field 
strengths) in different parts of the interacting system (Table~\ref{comp}). 
For all calculations we used maps of total and polarized radio 
intensity at 8.44~GHz. We adopted the face-on thickness of the synchrotron
emission to be 1~kpc (the typical nonthermal disk thickness, e.g. Condon 
\cite{condon}), a proton-to-electron energy ratio of 100 (Pacholczyk 
\cite{pacholczyk}), and a lower cutoff energy for relativistic protons of 
300\,MeV. The effective pathlengths were derived from the assumed inclination 
angle. We adopted an inclination of $73\degr$ for all regions associated 
with NGC~4038, $55\degr$ for NGC~4039 (both from LEDA database) and the mean 
of these values ($64\degr$) for the whole system and for regions between 
the galaxies. 

To account for all uncertainties due to a complex three-dimensional
geometry of the system we repeated the calculations for the whole system and 
the ``overlapping region'' with a mean inclination of $0\degr$. Also, for other 
regions where the value of inclination was uncertain, additional calculations 
were computed for the angle of $0\degr$. For the three regions discussed in 
Sect.~\ref{magneticridge} we performed calculations with inclinations 
determined from our models. All the results are presented in Table~\ref{comp}. 
The typical uncertainty in the calculated values is about 
25--30$\%$, including an uncertainty of a factor of two in
the nonthermal face-on disk thickness, the proton-to-electron energy ratio and 
in the lower proton energy cutoff.

For the whole system we averaged the radio intensities
within the area delineated by the intensity level of 0.3~mJy/b.a. 
(roughly 1\% of the peak intensity) and obtained a mean total 
magnetic field strength of $B_{\mathrm{tot}}=20\pm 5\,\mu$G. This is a high value, 
almost as high as in the starburst galaxy M~82 (Klein et al. \cite{klein88}). 
The mean regular field is B$_{\mathrm{reg}}=5\pm1\,\mu$G and yields a field 
regularity 0.25. Compared to other galaxies the {\bf mean} field regularity 
is very low (see Sect.~\ref{origin}).
A smaller face-on nonthermal disk thickness would even yield 
stronger magnetic fields. A thickness of 0.5~kpc, as used for NGC~4038/39 
by Hummel \& van der Hulst (\cite{hummel}), and an inclination of $64\degr$ 
yield a total field strength of $24\,\mu$G  for the whole system.

The numbers in Table~\ref{comp} show that the total magnetic fields in the 
individual galaxies without the ``overlapping region'' and in the ``overlapping 
region'' itself are about twice stronger than in normal spirals for 
which the mean value depending on the sample is 8--10$\,\mu$G (see Beck et 
al. \cite{beck96}). This field amplification must be a result of 
interaction.

Locally, the strongest magnetic fields of $B_{\mathrm{tot}}\simeq 30\,\mu$G are 
observed {\it outside} the galaxy disks, in the southern part of the 
``overlapping region'' (C and D in Fig.~\ref{sum}), a phenomenon which was 
never observed before in any galaxy system. This strong magnetic field is likely
caused by the interaction. In quiescent regions in the northwestern 
part of NGC~4038 magnetic fields resembles that of a normal spiral arm.
In region F, which is probably the least disturbed place of the relic spiral
arm located away from strong star-forming regions, the magnetic field 
reaches a value of $23\,\mu$G (for 1\,kpc disk thickness), comparable to that 
observed in the brightest parts of strong spiral arms in e.g. NGC~6946 
and NGC~1566 (Beck et al. \cite{beck96}). The most quiet ``interarm'' 
region in NGC~4038 (the last region presented in Table~\ref{comp}) shows 
$B_{\mathrm{tot}}\simeq 22\,\mu$G, much higher than for typical interarm 
regions in normal galaxies. However, this place is not far away from the 
central region of NGC~4038 so that some enhancement of star formation is 
suspected.

\subsection{The regularity and origin of magnetic fields in NGC~4038/39}
\label{origin}

Particular regions with different field strengths and structures can be 
associated with various phenomena in the gas dynamics in NGC~4038/39.
Regular magnetic fields in NGC~4038/39 are similar to that observed in 
normal spirals and are of order of 5$\,\mu$G. However, the field 
regularity is rather low (0.2--0.3 even in the ``interarm'' region).
A typical regularity of $0.5\pm0.1$ was found for normal galaxies 
(Buczilowski \& Beck \cite{buczilowski}, Beck et al. \cite{beck96}).
Our resolution relative to the galaxy size is better than in all
galaxies used in the quoted papers.
In such case we should expect even higher degree of field regularity 
($>0.5$) because of lower beam depolarization. This means that in the 
interacting system either strong amplification of turbulent fields
occurs (see below), or both turbulent and regular magnetic fields are 
generated, but the regular fields are much more tangled in the starburst 
regions than in normal spirals. Strong regular magnetic fields 
($\approx 8\,\mu$G) are observed close to NGC~4038 center and in its 
relic spiral arm (region F) but still with a low field regularity of 
0.3--0.4. In the northeastern ridge, far away from star-forming regions, 
the magnetic fields become more ordered and reach a strength of 
$\geq10\,\mu$G (region A), probably as the result of gas shearing motion 
in the base of the tidal tail.

In the northeastern part of the system (region A) the response of magnetic 
fields to the local gas flows is well visible. The magnetic pitch angle change 
from large values ($>40\degr$) in the western and northern part of NGC~4038 
to small ones in the eastern part, reflecting the alignment of regular 
magnetic field with the tidal tail, probably due to the gas stretching 
motions while a higher field regularity is suggestive for decaying gas 
turbulent motions. In the northern (region B) and middle part of the 
``overlapping region'' magnetic fields are also aligned with the shape 
of this region possibly adjusting to the gas flows.

The enhancement of total magnetic field in the ``overlapping region'' (C and D) to 
high values of $\simeq 30 \,\mu$G with coherence scale of about 1 kpc (visible 
in e.g. Fig.~\ref{sum}) could be produced by the fluctuation dynamo 
(Subramanian \cite{subramanian}) which may amplify magnetic fields liberated 
from the parent galaxies during the gravitational interaction. However, 
assuming a turbulent magnetic diffusivity of about 
$10^{27}\, \mathrm{cm}^2 \mathrm{s}^{-1}$ (an intermediate value between a 
typical value in the disk and in the corona in galaxies) the estimated 
magnetic diffusion time is about $3\times10^8$~yr, very close to the dynamic 
scale time of this region of about $2 \times 10^8$~yr (e.g. Mihos et al. \cite{mihos}). 
Thus, no amplification mechanism by any dynamo action is necessary to 
sustain the magnetic fields in this region if strong enough magnetic fields 
were already present in the disks. 

Instead we may assume isotropic compressional enhancement of magnetic 
fields in region C and D according to the approximate scaling with confining 
gas density as $B\propto\rho^{2/3}$, with $10\,\mu$G of the magnetic field 
strength before compression and the observed value of $30\,\mu$G after 
compression. Current gas density estimated from the total mass of gas 
approximated by the sum of molecular gas (from CO observations, Zhu et al. 
\cite{zhu}) and neutral $\ion{H}{i}$ gas (from Hibbard et al. \cite{hibbard}) 
yields a value of $10.7\,\mathrm{cm}^{-3}$. Using the scaling relation we get a gas 
density of $2.1\,\mathrm{cm}^{-3}$ before compression. This value is similar 
to gas densities in normal galaxies, e.g. in M31 where gas densities span 
a range of 1--6\,cm$^{-3}$ (Berkhuijsen et al. \cite{berkhuijsen}).
Hence, compressional enhancement in the ``overlapping region'' of magnetic 
fields pulled out from the parent disks is highly probable. Nevertheless some 
kind of fluctuation dynamo cannot be excluded, but this needs elaborate MHD 
modelling.

\subsection{Evolution of gas and magnetic field in NGC~4038/39}
\label{evolution}

The coupling of magnetic field with different gas phases shows more variety
in NGC4038/39 than in normal galaxies where density waves play an ordering 
role. The profiles of various species along three different 
directions (discussed in Sect.~\ref{thnth}) revealed different 
connections of magnetic fields with cold, warm and hot gas depending on 
the particular location and dominance of various physical processes. 
The intensity ratios between infrared, X-ray and radio emissions yielded 
additionally a crude estimate of the evolutionary stages of the selected 
places.

From all the above considerations we propose the following tentative scenario 
of the gas and magnetic field evolution in NGC~4038/39:
\begin{itemize}

\item[1.] Before the present encounter the northern galaxy was gas rich with 
a strong spiral magnetic field.

\item[2.] The current collision has pulled out the stellar tidal tails (in 
agreement with simulations by Englmaier, priv. comm., see also Englmaier et 
al. \cite{englmaier}). The NW part of ``inverted 9'' structure is possibly a 
relic of an unfolded 
spiral arm and  constitutes the base of an expanding gas-rich tail extending 
further to the south. The regular magnetic field remains aligned with the outer 
parts of this structure (region A). The magnetic field gets more ordered as 
the turbulence decays in the quiet, SE parts of this feature.

\item[3.] Another spiral arm of the original northern galaxy has possibly 
collided with that of the southern galaxy, giving rise to the dark cloud 
complex with a regular field aligned along the collision front (regions B, C and D). 
The magnetic field may be already strong in the northern, cool part of the 
complex (region B) but there is not yet strong star formation able to produce CR 
electrons. Some 5--10 Myr ago a strong star formation burst occurred along 
the NE boundary of the cloud (region E), efficiently destroying the regular magnetic 
field and filling this region with a pool of diffuse, hot gas.

\item[4.] A young star formation burst has occurred in the southern part of 
the cloud (region D), being still highly obscured in the optical domain, 
but visible in radio and infrared emission. It already managed to disrupt 
the regular magnetic field and to build up exceptionally strong total fields 
by compressional forces. At the southern edge of this clouds complex (region C), 
star formation is so young that radio intensity is still weak.

\end{itemize}

\section{Summary and conclusions}
\label{summary}

We present the first study of radio polarized and total emission
of a pair of spiral galaxies in the process of collision. The galaxies -- 
NGC~4038/39 -- were observed with the VLA in the C and D arrays at 
8.44~GHz, 4.86~GHz and 1.49~GHz. 

We show that the galaxies possess bright, substantially polarized radio disks 
with no dominant central sources. The thermal fraction is found to be about 
$50\%$ at 10.45~GHz, higher than in normal spirals. The derived magnetic fields 
in both galaxies are about two times stronger ($\simeq 20\,\mu$G) 
than in normal spirals probably as the result of interaction-enhanced 
star-formation and stronger turbulent motions. A similarly strong mean
magnetic field is also found along the interface between the two disks
known as the place of merger-driven extensive starburst. 
In comparison with other galaxies magnetic fields in the Antennae have lower 
regular components relative to the total one which is likely due to more 
strongly tangled fields in the vivid star-forming regions. 

Our radio data are compared with existing data in \ion{H}{i}, H$\alpha$, 
X-rays and in the infrared to study the physical conditions in various 
places of the merger and local interrelations of magnetic fields with 
different ISM phases. We distinguish several regions in the interacting 
system (depicted as A--F in Fig.~\ref{sum}) with the following results:
\begin{itemize}
\item [A.] The eastern highly polarized ridge where the gas-rich southern 
tidal tail starts. This region has no counterpart in any of star 
formation-dependent quantities like total radio, thermal, nonthermal or 
infrared intensity. It coincides with the \ion{H}{i} ridge which extends far 
out along the southern tidal tail. The radio spectrum is much steeper in this 
region and still steepens up to $\alpha >1.2$ towards the tail, which 
indicates the CR electron population aging as they move away from their 
sources in star-forming regions. The magnetic field is highly regular 
($B_{\mathrm{reg}}\simeq 10\,\mu$G) and stretched along the ridge, becoming 
even more ordered towards the tidal tail as turbulent motions decay, 
it also bends to become almost parallel to the sky plane.

\item[B.] The cold northern part of a dark cloud complex, constituting 
the ``overlapping region''. The radio total, thermal and nonthermal 
intensity is rather weak, indicating no hidden star formation. We 
note also that the whole region of the dark cloud complex shows a coherent 
magnetic field structure, probably tracing the line of collision between the 
arms of discussed merging spirals.

\item [C.] The hidden star-forming clump at the southern tip of the 
dark cloud complex. The clump is very strong in the infrared but 
deficient in the total radio intensity relative to the infrared one. Either 
the star formation burst occurs in rather weak magnetic field around or it 
is so young ($<10^6$~yr) that not enough CR electrons were produced.

\item[D.] The region of hidden star formation in the southern part of the 
dark cloud complex bright in total radio, thermal, 
nonthermal and infrared intensity. However, the H$\alpha$ peak is rather weak due 
to heavy extinction. The region is weakly polarized and most of the 
magnetic field is random reaching large total values of $B_{\mathrm{tot}}\simeq 
30\,\mu$G. This is probably the result of compression of the original fields 
pulled out with gas from parent galaxies. X-ray emission is relatively weak, 
possibly the starburst is still too young (i.e. $<10^7$~yr) to fill this 
volume with hot gas.

\item[E.] The X-ray ridge tracing an older ($>5\times 10^6$~yr) star formation burst 
leaving a pool of the hot gas and intermediate age star clusters. It has no 
counterpart in total radio intensity (the CR electron population faded away) 
and the regular magnetic field has been efficiently destroyed by star-forming 
activity.

\item[F.] The relic of the magnetic spiral of the gas-rich northern galaxy. 
The Faraday rotation measure distribution shows that this region has a 
complex three-dimensional structure.

\end{itemize}

Locally, the strongest magnetic fields of $\simeq 30\,\mu$G are present outside 
the galaxy disks, in the ``overlapping region'', what was never observed before 
in any galaxy system. We also note the large values of the magnetic pitch 
angle in the relic spiral arm of NGC~4038 and the unusual ``interarm'' region 
with disrupted regular fields. Thus the magnetic fields in the Antennae are in 
structure and properties very distinct from what is observed in normal 
spirals.

The presented regions (A--F) reveal different physical conditions and evolutionary 
stages and magnetic fields associations with cold, warm and hot gas, depending 
on the particular place and dominance of various physical processes. 
This makes the Antennae a key target for studying galactic gas dynamics, 
the properties of various ISM phases, star-forming processes and their 
interrelations with the magnetic field. 

We are, however, 
still far from fully understanding the dynamics of NGC~4038/39. Further 
works involving the numerical simulations including the magnetic field and 
high resolution observations of gas kinematics of the Antennae in cold,  
warm neutral, and in ionized gas are highly desirable.  Especially
interesting would be a model of gas and magnetic field evolution in the 
``overlapping region'' far from the galaxy disks.

\begin{acknowledgements}
The Authors wish to express their thanks to Dr E. Hummel for providing his 
original VLA data on NGC~4038/39. We are very indebted to Dr J. Hibbard  
(NRAO) for providing us his \ion{H}{i} data and B/K image, Dr B.C. Whitmore 
(STScI) for the HST images, Dr G. Fabbiano for the {\em Chandra} image and
Dr I.F. Mirabel (CEA) for the ISO data. We also thank to Drs P. Englmaier,
M. Krause and Prof. M. Urbanik for very helpful discussions. KCh is 
grateful to Prof. R. Wielebinski for making possible his visits to 
Max-Planck-Institut f\"ur Radioastronomie in Bonn where part of this work 
has been done. We wish to express our  particular thanks to the anonymous 
referee for his valuable remarks and his effort to improve our paper. We 
have made use of the LEDA database (http://leda.univ-lyon1.fr). This work 
was supported by a grant from the Polish Research Committee (KBN), grant 
no. PB249/P03/2001/21.
\end{acknowledgements}

\end{document}